\documentclass[final]{siamltex}
\usepackage[dvips]{graphicx}
\usepackage{amsfonts}
\usepackage{amssymb}
\usepackage{amsmath}
\usepackage{color}
\usepackage{subfigure}
\usepackage{epsfig,overpic}
\usepackage{latexsym}
\usepackage{tikz}
\usepackage{todonotes}
\usepackage{verbatim}
\usetikzlibrary{decorations.pathreplacing}
\newcommand{\lbl}[1]{\label{#1}}
\newtheorem{remark}{Remark}[section]
\def\mbf{\mathbf}
\def\div{\mathrm{\,div}\,}
\def\divG{\mathrm{\, div}_\Gamma}
\def\dx{\mathrm{d}x}
\def\ds{\mathrm{d}s}

\def\tr{\mathrm{tr}}
\def\nGa{\mathbf{n}_{\Gamma}}
\def\nOm{\mathbf{n}_{\partial\Omega}}
\def\bsigma{{\boldsymbol{\sigma}}}
\def\btau{{\boldsymbol{\tau}}}
\def\bn{\mathbf{n}}
\def\fS{\mathbf{f}_S}
\def\fL{\mathbf{f}_L}
\def\bu{\mathbf{u}}
\def\bD{\mathbf{D}}

\def\bv{\mathbf{v}}
\def\bPG{\mathbf{P}_\Gamma}
\def\bPS{\mathbf{P}_S}
\def\nablaG{\nabla_\Gamma}
\def\cT{\mathcal{T}}
\title{Finite element methods  for a class of continuum models for immiscible flows with moving contact lines\footnotemark[1]}%
\author{Arnold Reusken\thanks{Institut f\"ur
Geometrie und Praktische  Mathematik, RWTH Aachen University, D-52056 Aachen,
Germany; email: {\tt reusken@igpm.rwth-aachen.de}}  \and Xianmin Xu\thanks{LSEC, Institute of Computational
  Mathematics and Scientific/Engineering Computing, 
  NCMIS, AMSS, Chinese Academy of Sciences, Beijing 100190, China
e-mail: {\tt xmxu@lsec.cc.ac.cn}} \and Liang Zhang\thanks{Institut f\"ur
Geometrie und Praktische  Mathematik, RWTH Aachen University, D-52056 Aachen,
Germany; email: {\tt liang@igpm.rwth-aachen.de}}
}
\begin{document}
\maketitle

\renewcommand{\thefootnote}{\fnsymbol{footnote}}

%\footnotetext[1]{}
%\footnotetext[2]{The State Key Laboratory of Scientific and Engineering Computing,\\
%Institute of Computational Mathematics and Scientific/Engineering Computing, Chinese Academy of Sciences, Beijing 100080, China.
%({\tt xmxu@lsec.cc.ac.cn}).}

\renewcommand{\thefootnote}{\arabic{footnote}}
\begin{abstract}
In this paper we present a  finite element method (FEM) for two-phase incompressible flows with moving contact lines.
We use a sharp interface Navier-Stokes model for the bulk phase fluid dynamics. Surface tension forces, including Marangoni forces and viscous interfacial effects, are modeled.  
%The two-phase hydrodynamics is modelled by a sharp-interface two-phase Navier-Stokes equation. 
For describing the moving contact we consider a class of continuum models which contains several special cases known from the literature.
For the whole model, describing bulk fluid dynamics, surface tension forces and contact line forces, we derive a variational formulation and a corresponding energy estimate. 
For handling the evolving interface numerically the level-set technique is applied.  The discontinuous pressure is accurately approximated by using a stabilized extended finite element space (XFEM). We apply a Nitsche technique to weakly impose the Navier slip conditions on the solid wall. A unified approach for discretization of the (different types of) surface tension forces and contact line forces is introduced.
%It gives flexbility for general geometry than the strong imposition. 
The numerical methods are first validated for relatively simple test problems, namely a stationary spherical droplet in contact with a flat wall and a spherical  droplet on a flat wall that spreads or contracts to a stationary form. A further validation is done for a two-phase Couette flow with contact lines. To illustrate the robustness of our FEM we also present results of simulations for a problem with a curved contact wall and for a problem with more complicated contact line dynamics.
\end{abstract}


%\begin{AMS}
%41A60,49J45,76T10
%\end{AMS}

\pagestyle{myheadings}
\thispagestyle{plain}
\markboth{ }{ }
\section{Introduction}
\label{sec:intro}
Two-phase flows with moving contact lines (MCL) are very common in nature and industry, e.g. droplet spreading, coating flows and two-phase flows in porous media.
It is well-known that the standard no-slip boundary condition will cause non-physical infinite dissipation near the moving contact line \cite{huh1971}. 
%The modelling of moving contact line and related problems are very active area topics recently\cite{bonn2009,snoeijer2013,sui2014}. 
There are several ideas on how to resolve this dissipation singularity. 
A very popular approach is to use slip type boundary conditions close to the contact line \cite{cox1986}. 
Another possibility is to assume some thin layer of surface flow \cite{shikhmurzaev1993}. 
Yet another approach is to use a diffuse interface model coupled with a no-slip  boundary condition \cite{jacqmin2000,yue2010}. 
%It is found that the diffusion of the interface will introduce some effective slip of the contact line\cite{yue2010}. 
Besides these basic approaches there is a wide range of other MCL models available \cite{Qian03, ren2007, dupont2010, snoeijer2013, sui2014}.
For an efficient and accurate simulation of such MCL models,  computational approaches that are used for two-phase flows without contact lines have to be adapted to the MCL situation\cite{renardy2001numerical,spelt2005,dupont2010,gao2012,bao2012,sui2013efficient,sprittles2013finite,gao2014efficient,xu2014level,shen2015efficient}. 
In this paper we present a finite element method for the simulation of two-phase incompressible flows with moving contact lines. We only consider \emph{sharp} interface models. Instead of presenting our method for one of the many MCL models known in the literature, we  consider a \emph{class} of sharp interface continuum models.
This class is  characterized by constitutive laws for the bulk fluid stress tensor $\bsigma$, the interface stress tensor $\bsigma_\Gamma$ and ``effective wall and contact line forces'' $\fS$, $\fL$. 
These constitutive laws are taken from the literature. For $\bsigma$ we restrict to the standard Newtonian stress tensor. 
For $\bsigma_\Gamma$ we consider not only the ``clean interface model'' but also models that include Marangoni effects and viscous behavior (Boussinesq-Scriven) of the interface. 
For the wall and contact line forces  we consider models very similar to the GNBC \cite{Qian03,Qian06,gerbeau2009}.
The wall force $\fS$ corresponds to the Navier boundary condition and the contact line force $\fL$ results from a contact line model. In this paper we do not address the (important) topic of how to model the contact line dynamics in two-phase incompressible flows, i.e., we do not investigate which choices for the wall force $\fS$ or the contact line force $\fL$ are (most) appropriate for certain flow problems. Our goal is to present a finite element method which has good accuracy and robustness properties for a large class of  two-phase sharp interface models with moving contact lines.  

Special difficulties in this  class of flow problems are the following. The bulk fluid stress tensor $\bsigma$ is \emph{discontinuous} across the (moving) interface $\Gamma$. 
The surface stress tensor $\bsigma_\Gamma$ is \emph{localized} on the interface and depends on the \emph{curvature} of $\Gamma$.
Often $\Gamma$ is only implicitly known (level set technique) and an accurate  numerical approximation of $\bsigma_\Gamma$ is a difficult task. 
The contact line force $\fL$ is \emph{localized} on the moving contact line and in general depends on the \emph{dynamic contact angle}. 
An accurate  numerical approximation of these forces is hard to realize. Finally, the Navier boundary condition plays an important role in the model. 
An accurate handling of this boundary condition is very important. 
We treat numerical methods for dealing with these difficulties.

 The finite element solver is first validated for relatively simple test problems, namely a stationary spherical droplet in contact with a flat wall and a spherical  droplet on a flat wall that spreads or contracts to a stationary form. For both cases a comparison between analytical results and numerical simulation results is made.  A further validation is done for a two-phase Couette flow with contact lines. For this test case our results are compared with results from a molecular dynamics simulation \cite{Qian03}. To illustrate the robustness of our FEM we also present results of simulations for a problem with a curved contact wall and for a problem with more complicated contact line dynamics.

We outline the main contributions of this paper. We present a general class of sharp interface continuum models for two-phase incompressible flows with moving contact lines which contains several special cases known from the literature.
For such a  model, which describes the bulk fluid dynamics, different types of surface tension, wall and contact line forces, we derive a variational formulation and a corresponding energy estimate. This variational formulation  forms the basis for our finite element solver. This solver is based on several established techniques such as e.g., level set (combined with  reparametrization), a stabilized finite element discretization of the level set equation, continuous piecewise quadratics for the velocity approximation, implicit Euler time discretization for the fully coupled system. These methods can be found in the literature and are not addressed in this paper. We focus on the following three numerical techniques: 
\begin{itemize}
 \item A stabilized XFEM for the pressure, to deal with the discontinuity in $\bsigma$. This method has recently been introduced and analyzed in \cite{Hansbo14} for a stationary two-phase Stokes problem. We apply this method to two-phase incompressible flow problems with moving contact lines. As far as we know, the stabilized XFEM has not been used for this problem class, yet.
 \item A Nitsche technique for a flexible and accurate handling of the Navier boundary condition. This Nitsche method has recently been studied in \cite{urquiza2014weak} for treating the Navier boundary condition in a one-phase stationary Stokes problem. We are not aware of any literature in which this technique is applied to two-phase flows with (moving) contact lines. 
\item A  unified finite element discretization approach for  different types of surface tension and contact line forces.
\end{itemize}
As already noted above, our aim is not the comparison and/or validation of certain continuum models (e.g. GNBC) for contact line dynamics, but to present  finite element techniques that are accurate and robust for a large class of relevant models. 

The rest of the paper is organized as follows. In section~\ref{sectmodel}, we outline the physical background and introduce a class of models in strong formulation.
%The GNBC conditon is used to describe the movement of contact lines.
A corresponding variational formulation and an energy estimate for the solution of this variational problem are presented in section~\ref{sectvariational}. 
In section~\ref{sectLS} we very briefly treat the discretization of the level set equation and the reconstruction of the approximate interface. The three (new) numerical techniques mentioned above are explained in the sections~\ref{sectXFEM}, \ref{sect|Nitsche}, \ref{sectforces}. Results of numerical experiments are presented in section~\ref{sectExp}. In section~\ref{sectoutlook} we draw conclusions and give an outlook.
%\todo[inline]{Todo: a few literature are missing}
%There are many models for moving contact line problems.
%The diffuse interface model with no-slip boundary conditions will introduce some effective slip\cite{?,Yue}.  The Navier slip boundary condition near the
%moving contact lines also used to cure this paradox\cite{Cox,}. The generalized Navier Slip Boundary Condition(GNBC) is also introduced from molecular dynamics simulations and has been consistent with a diffuse interface model. A boundary condition for a sharp interface model is also introduced by Ren and E. In this paper, we consider the numerical method for a sharp interface model with generalized Navier slip boundary conditions. We start by a variational formula for the model. The variational formula is found also a weak formula for the sharp-interface model assuming microscopic contact angle to be Young's angle and Navier slip boundary conditions.
%
%It seems that an efficient numerical method should be given for the model.

\section{Physical background and a sharp-interface model in strong formulation} \label{sectmodel}

\begin{figure}[ht!]
\begin{center}
  \begin{tikzpicture}[scale=1.7, font=\footnotesize]
     %the slip plate
    \draw (0,0) -- (4,0) -- (5,2.5) -- (1,2.5)--cycle;
    \draw [decoration={border,segment length=0.9mm,amplitude=0.8mm,mirror,angle=45} , decorate] (0,0) -- (4,0) -- (5,2.5) -- (1,2.5)--cycle;
    \node at (0.4,0.3) { $\partial\Omega_S$};
    %droplet
    \draw [domain=0:180] plot ({2.5 +1.002*cos(\x)}, {1.25 +1.002*sin(\x)});
    \draw [blue](1.5,1.25) arc (180:360:1 and 0.4);
    \draw [blue,dashed] (3.5,1.25) arc (0:180:1 and 0.4);
    %slip boundary normal, the contact line nomal, and the droplet surface normal
    \draw [->]({2.5 +1*cos(-60)}, {1.25 +0.4*sin(-60)})--({3 +1*cos(-60)}, {1 +0.4*sin(-60)});
    \node at ({3.1 +1*cos(-60)}, {1.05 +0.4*sin(-60)}) {$\bn_{L}$};
    \node at ({2.5 +0.85*cos(-60)}, {1.25 +0.34*sin(-60)}) {$L$};
    \draw [dashed, ->] ({2.5 +1*cos(-60)}, {1.25 +0.4*sin(-60)})--({2.5 +1*cos(-60)}, {0.75 +0.4*sin(-60)});
    \node at ({2.5 +1*cos(-60)}, {0.65 +0.4*sin(-60)}) {$\bn_{S}$};
    \draw [->]({2.5 +1*cos(45)}, {1.25 +1*sin(45)})--({2.5 +1.5*cos(45)}, {1.25 +1.5*sin(45)});
    \node at ({2.6 +1.5*cos(45)}, {1.25 +1.3*sin(45)}) {$\mbf n_{\Gamma}$};
    \node at ({2.5 +0.8*cos(45)}, {1.25 +0.8*sin(45)}) {fluid 1};
    \node at ({2.5 +1.65*cos(45)}, {1.25 +1.65*sin(45)}) {fluid 2};
    %contact angle
    \draw  [domain=0:84] plot ({1.5 +0.15*cos(\x)}, {1.25 +0.15*sin(\x)});
    \draw  [dashed] (1, 1.25)--(2,1.25);
    \node at (1.7,1.35){ $\theta$};
 
    %\node at (2,1.5) {$\Omega 2$};
    %\draw (1,0.1)--(1.1,0.1)--(1.1,0);
    %\node at (1.6,0.15) [font=\tiny]{$\theta=90^{\circ}$};
  \end{tikzpicture}
  \caption{A liquid droplet with a contact line.\label{fig:cl}}
\end{center}
%   \label{fig:cl}
\end{figure}
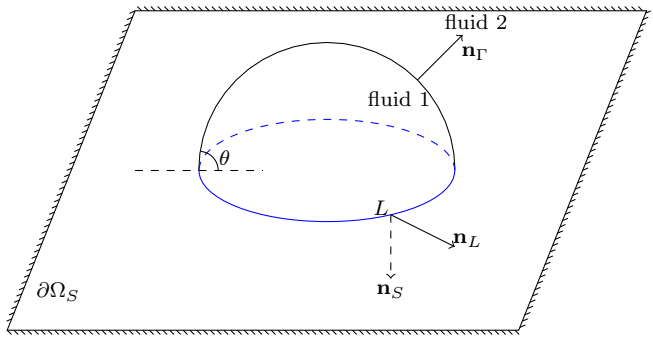

We introduce a sharp-interface model for immiscible two-phase flow with a moving contact line. We consider  an immiscible two-phase flow in a polygonal domain $\Omega \subset \mathbb{R}^3$.
The time-dependent domains of each fluid are denoted by $\Omega_1:=\Omega_1(t)$ and $\Omega_2:=\Omega_2(t)$.
The evolving sharp interface between the two fluids is denoted by $\Gamma(t) := \partial \Omega_1 \cap \partial \Omega_2 $.  We assume that part of $\partial \Omega$ consists of a plane, denoted by $\partial \Omega_S$ (``sliding wall'') that is in contact with $\Omega_1$, cf. Figure~\ref{fig:cl}. The contact line is denoted by $L$ and the normals on $\Gamma$ and $\partial \Omega_S$ are denoted by $\bn_\Gamma$ and $\bn_S$, respectively. The normal to $L$ lying in  $\partial \Omega_S$ is denoted by $\bn_L$. 

A continuum model for the fluid dynamics is based on mass and momentum conservation combined with constitutive laws for stress tensors  $\bsigma$, $\bsigma_\Gamma \in \mathbb{R}^{3 \times 3}$ and (virtual) forces $\fS, \fL \in \mathbb{R}^3$  that describe (effective) forces in the bulk fluids, at the interface, the wall $\partial \Omega_S$ and the contact line, respectively. We only outline the main ideas. For more detailed treatments we refer to the literature, e.g., \cite{Slattery,Gurtin,Sagis}. We restrict to incompressible bulk fluids, i.e. $\div \bu =0$ in $\Omega_i$, $i=1,2$, with $\bu$ the velocity. \\[1ex]
{\bf Bulk fluid stress tensor.} For the bulk fluid  stress tensor $\bsigma=\bsigma_i(x,t)$, $i=1,2$, we only consider the standard Newtonian model
\begin{equation} \label{Newton} 
\bsigma_i:=-p\mbf I +\mu_i \mbf D(\bu), \quad \mathbf D (\mbf u):= \nabla \mbf u+\nabla \mbf u^T,\quad i=1,2,
\end{equation}
with $p$ the pressure and $\mu_i >0$ the constant viscosity of fluid $i$. \\[1ex]
{\bf Interface stress tensor}. The interface stress tensor $\bsigma_\Gamma$  acts only in planes tangential to $\Gamma$. Hence $\bsigma_\Gamma$ has to fulfill the condition $\bPG\bsigma_\Gamma \bPG= \bsigma_\Gamma$, with the orthogonal projection $\bPG:=\mbf I - \bn_\Gamma \bn_\Gamma^T$. Let $\gamma$ be a small connected subset of $\Gamma$ with boundary $\partial \gamma$ and a normal $n_\gamma$ which is orthogonal to $\partial \gamma$ and tangential to $\Gamma$, hence $\bPG n_\gamma = n_\gamma$. The stress tensor $\bsigma_\Gamma$ models the contact force in the interface $\Gamma$, i.e., $\bsigma_\Gamma n_\gamma$ is the force density on $\partial \gamma$ (and thus a force per unit of length). 
This interface (line) contact force induces a surface force in $\Gamma$ given by $\divG \bsigma_\Gamma$.  A standard  constitutive law is given by
\[ \bsigma_\Gamma^0:=\tau \bPG,
 \]
with a (variable) surface tension coefficient $\tau >0$, which may be a function of $x \in \Gamma$. If $\tau$ is constant this model corresponds to a  ``clean interface''. Models with a variable $\tau$, e.g., $\tau$ depending on a local surfactant concentration, lead to so-called Marangoni effects. A standard constitutive law to describe  interfacial viscous forces is the following Boussinesq-Scriven model:
\begin{equation} \label{BS}
\begin{split}
 \bsigma_\Gamma^{BS} & =\bsigma_\Gamma^0 + \bsigma_\Gamma^{visc} \\
  \bsigma_\Gamma^{visc} &:= \lambda_\Gamma (\divG \bu) \bPG + \mu_\Gamma \mbf D_\Gamma(\bu), ~~\mbf D_\Gamma(\bu)=\bPG\big(\nabla_\Gamma \bu + (\nabla_\Gamma \bu)^T\big)\bPG 
\end{split} \end{equation}
with $\lambda_\Gamma > - \mu_\Gamma$ and $\mu_\Gamma>0$ the interface dilatational viscosity  and interface shear viscosity, cf.~\cite{Slattery}. Note the similarity to the bulk tensor model in \eqref{Newton}. In the Newtonian stress tensor \eqref{Newton} the term $\div \bu$ does not occur, due to the incompressibility assumption $\div \bu=0$. In general $\div \bu=0$ does not imply $\divG \bu=0$. If the interface has visco-elastic behavior, the stress tensor $\bsigma_\Gamma$ has to be adapted to this. 
The development of such visco-elastic surface stress tensors is an active and relatively new research area \cite{Sagis}.   In the remainder of this paper we only consider  interface stress tensors
\[
  \bsigma_\Gamma \in \{ \bsigma_\Gamma^0, \bsigma_\Gamma^{BS} \}.
\]
We expect, however, that the methods that are treated can also be applied (with minor modifications) to problems with other interface stress tensors.\\[1ex]
{\bf Contact line  model.} There is extensive literature on the modeling of dynamic contact lines. We outline a class of models used in many papers, e.g., \cite{Buscaglia2011,Qian03,Qian06,ren2007,Ren2011,Ren2010,Shikhmurzaev97}. Typically, for a sharp interface continuum model one uses effective (virtual) dissipative forces on the wall $\partial \Omega_S$ and at the contact line $L$. Let $\bn_S$ and $\bn_L$ be the normals as illustrated in Figure~\ref{fig:cl} and $\bPS=\mbf I - \bn_S \bn_S^T$ the orthogonal projection on the plane that contains   $\partial \Omega_S$. 

At the contact line there is a wall surface tension force $(\tau_2 - \tau_1)\bn_L$, with $\tau_1$ and $\tau_2$ the surface tension coefficient for wall-fluid 1 and wall-fluid 2, respectively. Using Young's equation $\tau \cos \theta_e= \tau_2 - \tau_1$, with $\theta_e$ the static contact angle, this wall surface tension force can be represented as $\tau \cos \theta_e \bn_L$. 
We assume that the wall effects can be modeled using effective (friction) force densities. More specifically we assume ``wall effect'' forces of the form
\begin{align}
  \fS & = - \beta_S \bPS \bu \quad \text{on}~~\partial \Omega_S  \label{FS}\\
 \fL & = -\beta_L (\bu \cdot \bn_L)\bn_L +  \tau \cos \theta_e \bn_L \quad  \text{on}~~ L,  \label{FL}
\end{align}
where  $\beta_S,\beta_L \geq 0$ may depend on the flow variables and on $x \in \partial \Omega_S$. Note that the second term on the right hand-side in \eqref{FL} is the wall surface tension force and the first term models dissipative friction effects. In the remainder we restrict to the linear case, in which $\beta_S,\beta_L \geq 0$ are assumed to be given functions, depending only on  $x \in \partial \Omega_S$ (e.g., constants). The force $\fS$ is an area force density, whereas $\fL$ is a line force density. Note that both forces are in the plane that contains $\partial \Omega_S$, i.e. $\bPS \fS= \fS$, $ \bPS \fL= \fL$. For a motivation and validation of this ansatz, using  thermodynamic principles, energy arguments and molecular dynamics simulations, we refer to the literature \cite{Buscaglia2011,Qian03,Qian06,ren2007,Ren2011,Ren2010,Shikhmurzaev97}. For  the contact line force $ \fL$ it is convenient to introduce a corresponding contact line stress tensor  $\bsigma_L \in \mathbb{R}^{3 \times 3}$:
\begin{equation} \label{GNBC}
 \fL = \bsigma_L \bn_L, \quad  \bsigma_L = \big[-\beta_L (\bu \cdot \bn_L) +  \tau \cos \theta_e \big] \mbf I. 
\end{equation}
 \\[1ex]
{\bf Boundary conditions.}   On $\partial \Omega_S$ we assume the no-penetration condition $\bu \cdot \bn_S  = 0$. This can also be written as $ (\mbf I-\bPS)\bu  = 0$ on  $\partial \Omega_S$.   On the remaining part of the boundary $\partial \Omega_D:= \partial \Omega \setminus \partial \Omega_S$ we assume no-slip boundary conditions: $\bu = 0$ on  $\partial \Omega_D$. The results of this paper apply, with minor modifications, if on $\partial \Omega_D$ we consider inhomogeneous Dirichlet boundary conditions $\bu=\bu_D$ instead  homogeneous ones, or if on part of $\partial \Omega \setminus \partial \Omega_S$ we have a natural boundary condition $ (-p\mbf I+\mu \mbf D(\mbf u))\nOm= -p_{ext}\nOm$. 
\\[1ex]
{\bf Model in strong formulation.} From mass conservation and the incompressibility assumption we obtain the equation $\div \bu =0 $ in $\Omega_i$. 
With respect to momentum conservation, we assume that the only forces in the system are  an external volume force (gravity)  denoted by $\mbf g$, viscous contact forces in the bulk phases modeled by the stress tensor $\bsigma$, interface contact forces modeled by $\bsigma_\Gamma$ and the wall forces given in \eqref{FS}, \eqref{FL}.   Momentum conservation in a material volume contained in $\Omega_i$ results in 
\[
  \rho_i \big( \frac{\partial \bu}{\partial t} + \bu \cdot \nabla \bu\big) = \div \bsigma_i + \rho_i \mbf g \quad \text{in }~\Omega_i,
\]
where $\rho_i >0$ denotes the constant density of the fluid in $\Omega_i$.  Momentum conservation in a material volume that is intersected by $\Gamma$ (but not by $\partial \Omega_S$)  yields the force balance
\[
  [\bsigma \bn_\Gamma] = \divG \bsigma_\Gamma \quad \text{on}~~\Gamma,
\]
with $[\cdot]$ the usual jump operator across $\Gamma$. Momentum conservation in a material volume with part of its boundary on $\partial \Omega_S$ (and no intersection with $\Gamma$) implies the force balance condition 
 \begin{align} 
 -\beta_S \bPS \bu &= \bPS \bsigma \bn_S =\mu \bPS\mathbf D (\mbf u) \bn_S \quad \text{on}~~\partial \Omega_S.\label{NavierBC2}
\end{align}
Note that this condition, combined with the no-penetration condition $\bu \cdot \bn_S  = 0$ is the usual Navier slip boundary condition on $\partial \Omega_S$.
Similarly, using a material volume the boundary of which is intersected by the contact line, momentum conservation implies the force balance condition 
\[
   \bsigma_L \bn_L= \bPS \bsigma_\Gamma \btau_L \quad \text{on}~~L,
\]
where $\btau_L= \frac{\bPG \bn_S}{ \|\bPG \bn_S\|}$ is the normal to $L$ that is tangential to $\Gamma$.

To obtain a closed model, we need further conditions. From  the assumption that there is no slip at the interface and both fluids are viscous we get $[\bu]=0$ on $\Gamma$. The immiscibility assumption leads to $V_\Gamma= \bu \cdot \bn_\Gamma$, where $V_\Gamma$ denotes the normal velocity of the interface. Summarizing we obtain the following \emph{two-phase sharp interface model} 
\begin{align}
 & \left\{
\begin{array}{rl}
 \rho_i(\frac{\partial \mbf{u}}{\partial t}+\mbf u\cdot\nabla\mbf u) & = \div \bsigma_i +\rho \mbf{g} \\
 \div \mbf u & =0, 
\end{array} \right.\lbl{e:NSsys}  \quad \text{in} ~~\Omega_i(t),~i=1,2, \\
&  [\bsigma \bn_\Gamma ]  = \divG \bsigma_\Gamma,~~V_\Gamma=\mbf u\cdot\nGa,\quad [\mbf u] =0 ~~ \hbox{on } \Gamma(t),\label{e:Gamma}\\
 &  \bsigma_L \bn_L= \bPS \bsigma_\Gamma \btau_L \quad \text{on}~~L(t), \label{e:Cline} \\
&  \left\{ \begin{array}{rl} (\mbf I-\bPS)\bu & = 0    \\
 -\beta_S \bPS \bu &=  \bPS \bsigma \bn_S 
 \end{array} \right. \quad \text{on}~~\partial \Omega_S,\label{e:NavierBC} \\
& \mbf u= 0 \quad \hbox{on }\partial \Omega_D. \label{DirichletBC}
\end{align}
For well-posedness we in addition need suitable initial conditions for the velocity $\bu(x,0)$ and the initial interface $\Gamma(0)$. We are interested in the solution $\bu(x,t)$, $p(x,t)$ (and $\Gamma(t)$) of this model, for $(x,t) \in \Omega \times [0,T]$.
\begin{remark} \label{remmodels} \rm  We emphasize that there is still considerable freedom in the choice of the stress tensor $\bsigma_\Gamma$ and the wall forces $\fS$, $\fL$. In this sense \eqref{e:NSsys}-\eqref{DirichletBC} forms a \emph{class} of sharp interface models. For special choices we relate the model above to models studied in the literature. 
For the choice $\bsigma_\Gamma= \bsigma_\Gamma^0$ we have, cf.~\cite{GrossReuskenBook},
\[ \divG \bsigma_\Gamma= \div_\Gamma (\tau \bPG)= - \tau \kappa \bn_\Gamma +\nabla_\Gamma \tau
\]
with $\kappa$ the curvature of $\Gamma$. Hence,  if $\tau$ is constant  the stress tensor jump condition in \eqref{e:Gamma} takes the familiar form $[\bsigma \bn_\Gamma ]=- \tau \kappa \bn_\Gamma$.  For this choice $\bsigma_\Gamma=\bsigma_\Gamma^0= \tau \bPG$ we have $\bPS \bsigma_\Gamma \btau_L = \tau \bPS \bPG \btau_L=\tau \bPS \btau_L= \tau \cos \theta \, \bn_L$ and thus the force balance in \eqref{e:Cline} takes the form
\begin{equation} \label{l1}
  \beta_L U_L= \tau (\cos \theta_e - \cos \theta) \quad \text{on}~~L(t)
\end{equation}
with  $U_L:=\bu \cdot \bn_L$ the contact line normal velocity. For $\beta_L >0$ this relates the  contact line normal velocity to the \emph{uncompensated Young stress} or \emph{out-of-balance interfacial tension} $ \tau (\cos \theta_e - \cos \theta)$. Such a relation is studied in  the literature on dynamic contact line modeling, e.g.  \cite{Buscaglia2011,ren2007,Ren2010,Ren2011}.
 For $\beta_L=0$ the relation  in  \eqref{l1} combined with the Navier slip boundary condition in \eqref{e:NavierBC} form the so-called generalized Navier boundary condition (GNBC), which is treated in e.g., \cite{gerbeau2009,Qian03,Qian06}.
\end{remark}
\ \\

In this paper we do not address the (important) topic of modeling the contact line dynamics in two-phase incompressible flows, i.e., we do not study which choices for the wall forces $\fS$, $\fL$ are (most) appropriate for certain flow problems. The main goal is to present a finite element method which has good efficiency and robustness properties for the general two-phase sharp interface model  \eqref{e:NSsys}-\eqref{DirichletBC}. 

\section{Variational formulation} \label{sectvariational}

In this section
% we introduce a variational formula for the two-phase flow model, which is a basis for the finite element approximation. 
%An energy decay property is proved for the variational problem.
we introduce a variational formulation of the model \eqref{e:NSsys}-\eqref{DirichletBC} introduced above. 
This variational formulation forms the basis for the finite element discretization treated in the sections~\ref{sectXFEM}-\ref{sectFull}.
We introduce function spaces
\begin{align*}
\mbf X & :=\{\mbf v\in (H^1(\Omega))^3 : \mbf v=\mbf 0 \hbox{ on }\partial\Omega_D,~(\mbf I - \bPS)\bv =0\hbox{ on }\partial\Omega_S \},\\
Q & :=\{q\in L^2(\Omega):\int_{\Omega}q\, \dx =0 \, \}.
\end{align*}
Note that for $\bu \in \mbf X$, the condition $[\bu]=0$ in \eqref{e:Gamma} and the boundary conditions $(\mbf I- \bPS)\bu=0$ (no-penetration) in \eqref{e:NavierBC} and $\bu=0$ (Dirichlet BC) in \eqref{DirichletBC} are satisfied by definition of the space $\mbf X$ (i.e., essential boundary conditions).
It is well-known that the force balance $-\beta_S \bPS \bu =  \bPS \bsigma \bn_S $ in the Navier boundary condition \eqref{e:NavierBC} can be treated as a natural boundary condition in the variational setting. We will see that the same holds for the force balances $ [\bsigma \bn_\Gamma ]  = \div_\Gamma \bsigma_\Gamma$ (on $\Gamma$) and $ \bsigma_L \bn_L= \bPS \bsigma_\Gamma \btau_L$ (on $L$).

It is instructive to give a derivation of the variational formulation. We multiply the Navier-Stokes equation in \eqref{e:NSsys} by a sufficiently smooth test vector function $\bv$ with $\bv=0$ on $\partial \Omega_D$ and integrate over $\Omega_i$. We do the integration by parts
\[
 \int_{\Omega_i} \div \bsigma_i \cdot \bv \, \dx = - \frac12 \int_{\Omega_i} \mu_i \bD(\bu): \bD (\bv) \, \dx + \int_{\Omega_i} p \div \bv \, \dx +\int_{\partial \Omega_i} \bsigma_i \bn_{\Omega_i} \cdot \bv \, \ds.
\]
 The boundary $\partial \Omega_i$ is split into three parts $\partial \Omega_i \cap \partial \Omega_D$, $\partial \Omega_i \cap \partial \Omega_S$ and $\partial \Omega_i \cap \Gamma= \Gamma$. We have either $\partial \Omega_i \cap \partial \Omega_D = \emptyset$ or $\bv =0$ on $\partial \Omega_i \cap \partial \Omega_D$. Hence $\int_{\partial \Omega_i \cap\partial \Omega_D } \bsigma_i \bn_{\Omega_i} \cdot \bv \, \ds=0$. On $ \partial \Omega_S$ we write $\bv = \bPS \bv + (\bv \cdot \bn_S )\bn_S$, and using the force balance in \eqref{e:NavierBC} we get
\begin{align} 
 & \sum_{i=1}^2 \int_{\partial \Omega_i\cap \partial \Omega_S} \bsigma_i \bn_{\Omega_i} \cdot \bv \, \ds =\int_{\partial \Omega_S} \bPS \bsigma \bn_{S} \cdot \bPS \bv \, \ds + \int_{\partial \Omega_S} (\bn_S \cdot \bsigma \bn_S)( \bv \cdot \bn_S) \, \ds  \nonumber \\
 & =- \int_{ \partial \Omega_S} \beta_S \bPS \bu \cdot \bPS \bv \, \ds + \int_{\partial \Omega_S} (\bn_S \cdot \bsigma \bn_S)(\bv \cdot \bn_S ) \, \ds. \label{force1}
\end{align}
We now restrict to $\bv \in \mbf X$, hence $\bv \cdot \bn_S= 0 $ on $\partial \Omega_S$. Thus the second term on the right hand-side vanishes and we get
\begin{equation} \label{parti1}
 \sum_{i=1}^2 \int_{\partial \Omega_i\cap \partial \Omega_S} \bsigma_i \bn_{\Omega_i} \cdot \bv \, \ds=- \int_{ \partial \Omega_S} \beta_S \bPS \bu \cdot \bPS \bv \, \ds, \quad \bv \in \mbf X.
\end{equation}
For the $\partial \Omega_i \cap \Gamma$ boundary part we get, using the force balance in \eqref{e:Gamma},
\[ 
 \sum_{i=1}^2 \int_{\partial \Omega_i\cap \Gamma} \bsigma_i \bn_{\Omega_i} \cdot \bv \, \ds = \int_{\Gamma} [\bsigma \bn_\Gamma] \cdot \bv \, \ds = \int_\Gamma \divG \bsigma_\Gamma \cdot \bv \, \ds.
\]
Combining these results we obtain
\begin{equation} \label{eq11}
 \begin{split}
  \sum_{i=1}^2 \int_{\Omega_i} \div \bsigma_i \cdot \bv \, \dx &= -\frac12 \int_{\Omega}\mu \bD(\bu):\bD(\bv) \, \dx + \int_{\Omega} p \div \bv \, \dx \\
 &~~- \int_{ \partial \Omega_S} \beta_S \bPS \bu \cdot \bPS \bv \, \ds + \int_\Gamma \divG \bsigma_\Gamma \cdot \bv \, \ds, \quad \bv \in \mbf X.
 \end{split}
\end{equation}
Note that for the case $\bsigma_\Gamma= \bsigma_\Gamma^{BS}$ we need more smoothness than only $\bu \in \mbf X$, since $\bsigma_\Gamma$ contains a term $\nabla_\Gamma \bu$. We need sufficient smoothness of $\bu$ such that $(\nabla_\Gamma \bu)_{i,j} \in L^2(\Gamma) $. 
\\
In the derivation above, by using the integration by parts,  the force balances in \eqref{e:NavierBC} and \eqref{e:Gamma} are ``included'' in the bilinear form on the right hand-side in \eqref{eq11}. We treat the line force balance in \eqref{e:Cline} in the same way, namely by applying integration by parts to the term $\int_\Gamma \divG \bsigma_\Gamma \cdot \bv \, \ds$ that occurs in \eqref{eq11}. For any sufficiently smooth vector function $\bv$ the identity 
\begin{equation} \label{partial2}
 \int_\Gamma \divG \bsigma_\Gamma \cdot \bv \, \ds= - \int_\Gamma \bsigma_\Gamma: \nabla_\Gamma \bv \, \ds + \int_L \bsigma_\Gamma \btau_L \cdot \bv \, \ds
\end{equation}
holds. Using $\bv = \bPS \bv + (\bv \cdot \bn_S)\bn_S$ and the contact line force balance in \eqref{e:Cline} we get
\begin{align}
&  \int_L \bsigma_\Gamma \btau_L \cdot \bv \, \ds =\int_L \bPS \bsigma_\Gamma \btau_L \cdot \bPS \bv \, \ds +\int_L (\bn_S \cdot \bsigma_\Gamma \btau_L)( \bv \cdot \bn_S) \, \ds  \nonumber \\
&= \int_L \bsigma_L \bn_L \cdot \bPS \bv \, \ds +\int_L (\bn_S \cdot \bsigma_\Gamma \btau_L)( \bv \cdot \bn_S) \, \ds  \nonumber \\
 &= - \int_L \beta_L \bu \cdot \bn_L \, \bv \cdot \bn_L + \tau \cos \theta_e \, \bv \cdot \bn_L \, \ds +\int_L (\bn_S \cdot \bsigma_\Gamma \btau_L)( \bv \cdot \bn_S) \, \ds. \label{ref}
\end{align}
For $\bv \in  \mbf X$ we have $ \bv \cdot \bn_S =0$ and thus the third term in \eqref{ref} vanishes. Hence we get
\begin{equation} \label{parti2}
  \int_\Gamma \divG \bsigma_\Gamma \cdot \bv \, \ds= - \int_\Gamma \bsigma_\Gamma: \nabla_\Gamma \bv \, \ds - \int_L\beta_L \bu \cdot \bn_L \, \bv \cdot \bn_L \, \ds + \cos \theta_e \int_L \tau \bv \cdot \bn_L \, \ds
\end{equation}
in which the  line force balance \eqref{e:Cline} is now ``included''. 
Again, as noted above, we need more regularity than only $\bu, \bv \in \mbf X$ to guarantee that the quantities used in \eqref{parti2} are well-defined.
\\
For a compact representation of the variational problem we introduce bi- and trilinear forms and linear functionals:
\begin{align}
m(\mbf u,\mbf v)&:=\int_{\Omega} \rho \mbf u\cdot\mbf v\, \dx\nonumber,\\
a(\mbf u,\mbf v)&:=\frac{1}{2}\int_{\Omega} \mu\mbf D(\mbf u):\mbf D(\mbf v)\, \dx +\int_{\partial\Omega_S}\beta_S \bPS \mbf u\cdot \bPS \mbf v\, \ds + \int_{L} \beta_L \bu\cdot \bn_L \, \bv \cdot \bn_L \,\ds, \nonumber\\
b(\mbf v, q) &:=-\int_{\Omega}(\div \mbf v) q\, \dx, \nonumber\\
c(\mbf w;\mbf u,\mbf v)& :=\int_{\Omega}  \rho (\mbf w \cdot\nabla)\mbf u\cdot\mbf v\, \dx,\nonumber \\
f_{ext}(\mbf v) &:= \int_{\Omega}\rho \mbf{g}\cdot\mbf v\, \dx, \nonumber\\
f_{\Gamma}(\bu, \mbf v)&:=-\int_{\Gamma} \bsigma_\Gamma:\nabla_{\Gamma} \mbf v\, \ds, \nonumber\\
f_{L}(\mbf v) &:=\cos \theta_e \int_{L}\tau \bv \cdot \bn_L \, \ds\nonumber.
\end{align}
\begin{remark} \rm We comment on the surface tension force functional $f_{\Gamma}(\bu, \mbf v)$. First we consider $\bsigma_\Gamma= \bsigma_\Gamma^0= \tau \bPG$. Let ${\rm id}_\Gamma$ be the identity on $\Gamma$, i.e. ${\rm id}_\Gamma (x)=x$ for $x \in \Gamma$. From $\nabla {\rm id}_\Gamma= \mbf I$ and $\nabla_\Gamma =\bPG\nabla$ we obtain the representation
\[
  f_{\Gamma}(\bu, \mbf v)=-\int_{\Gamma} \tau \nabla_\Gamma {\rm id}_\Gamma:\nabla_{\Gamma} \mbf v\ds =:f_{\Gamma}^0(\mbf v), 
\]
which is often used in the literature, e.g. \cite{Baensch05,Deckelnick05,Gross06,GrossReuskenBook,Ausus2010,Hysing06,Saksono06}. A nice property of this surface tension representation is that curvature approximations are not (explicitly) needed. Note that in this case $ f_{\Gamma}$ depends only on $\bv$, hence it is a linear functional. For the Boussinesq-Scriven surface tension tensor we get, cf.~\eqref{BS},
\begin{align*}
 f_{\Gamma}(\bu, \mbf v) & =-\int_{\Gamma} \bsigma_\Gamma^0:\nabla_{\Gamma} \mbf v\,\ds- \int_{\Gamma} \bsigma_\Gamma^{visc}:\nabla_{\Gamma} \mbf v\, \ds \\
 &= f_{\Gamma}^0(\mbf v)- \lambda_\Gamma \int_{\Gamma} (\divG \bu ) \bPG :\nabla_{\Gamma} \bv \, \ds - \mu_\Gamma \int_{\Gamma}  \bD_\Gamma(\bu) :\nabla_{\Gamma}\bv \, \ds. 
\end{align*}
This bilinear form is used in \cite{ReuskenZhang}  for the numerical simulation of a two-phase flow problem with a viscous interface force (but without a contact line). Note that the viscous part is a bilinear form, i.e. depends  not only on $\bv$ but also on $\bu$. In the solution process this part is shifted to the left hand-side. 
\end{remark}
\ \\[1ex]
Although not explicit in the notation, the  bi- and trilinear forms and functionals depend on $t$, due to $\Gamma=\Gamma(t),~ L=L(t),~\rho=\rho(x,t),~\mu=\mu(x,t)$. 
For any given $t \in [0,T]$ let $\mbf X_t$ be a (possibly large) dense subspace of $\mbf X_0$ such that the  bi- and trilinear forms and functionals defined above are well-defined and continuous on $\mbf X_t$. The derivation above, cf. \eqref{eq11} and \eqref{parti2}, leads to the following variational formulation.
%Let $\bu=\bu(\cdot,t)$, $p=p(\cdot,t)$ be a (sufficiently smooth) solution of the strong formulation \eqref{e:NSsys}-\eqref{DirichletBC}. Then this pair also solves the variational equation:
Find $\mbf u = \mbf u(\cdot, t) \in \mbf X_t, p = p(\cdot, t) \in Q$ such that for (almost all) $t \in [0, T]$
 \begin{equation}
 \left\{
 \begin{array}{ll}
 m(\partial_t\mbf u,\mbf v)+c(\mbf u;\mbf u,\mbf v)+a(\mbf u,\mbf v)+b(\mbf v, p)=f_{ext}(\mbf v)+f_{\Gamma}(\bu,\mbf v) +f_{L}(\mbf v) 
  \\
b(\mbf u, q) =0
 \end{array}
\right.\lbl{e:vari1}
\end{equation}
for all $\bv \in \mbf X_t$, $q \in Q$. 
The derivation above, combined with standard variational arguments yields that a sufficiently smooth solution pair $(\bu,p)$ of the variational problem \eqref{e:vari1} is a solution of the strong formulation  \eqref{e:NSsys}-\eqref{DirichletBC}. In this sense the variational problem \eqref{e:vari1} is consistent with the model \eqref{e:NSsys}-\eqref{DirichletBC}. 
An interesting property of this formulation is that both \emph{curvature approximations and contact angle approximations are not needed}. 
In the remainder of this paper we consider the variational problem \eqref{e:vari1}. For the special case $\bsigma_\Gamma= \bsigma_\Gamma^0$ with a constant $\tau$, a numerical discretization method of this variational problem based on an ALE approach is presented  in \cite{ganesan2009,ganesan2013}. 
In the sections~\ref{sectLS}-\ref{sectFull} we present an Eulerian finite element method for the discretization of \eqref{e:vari1}. 
In the next subsection, we first derive an energy decay   property.

\subsection{Energy estimate}
An energy decay property of the weak formulation~\eqref{e:vari1} is shown in this subsection. We assume that the surface tension coefficients $\tau,\, \tau_1,\, \tau_2$ are constant. We introduce some notation. The wall boundary part  in contact with fluid $i$ is denoted by $\Gamma_S^{(i)}=\partial \Omega_i \cap \partial \Omega_S$.  The area of the interfaces $\Gamma(t)$ and $\Gamma_S^{(i)}(t)$ is denoted by $|\Gamma(t)|$ and $|\Gamma_S^{(i)}(t)|$. The norm $\|\cdot\|$ denotes the Euclidean vector norm.   We define the energy 
\[
 E_\bu(t)= \frac12 \int_\Omega \rho \|\bu\|^2 \, dx + \tau |\Gamma(t)| + \tau_1 |\Gamma_S^{(1)}(t)| + \tau_2 |\Gamma_S^{(2)}(t)|,
\]
which is the sum of kinetic and surface energies. 
\begin{lemma}
We assume $\mbf g=0$, i.e., $f_{ext}(\bv)=0$. We consider $\bsigma_\Gamma= \bsigma_\Gamma^{BS}$. Let $(\mbf u,p)$ be a solution of \eqref{e:vari1} for $t\in [0,T]$. The following  holds:
\begin{align}
 \frac{d}{dt} E_\bu(t) & = - \Big[ \frac12 \int_\Omega \mu \tr \big( \bD (\bu)^2\big) \, \dx + \lambda_\Gamma \int_\Gamma \tr (\nablaG \bu)^2 \, \ds + \frac12 \mu_\Gamma \int_{\Gamma} 
 \tr \big(\bD_\Gamma(\bu)^2\big) \, \ds  \nonumber \\ &  \quad +\int_{\partial \Omega_S} \beta_S \|\bPS \bu\|^2 \, \ds + \int_L \beta_L (\bu \cdot \bn_L)^2 \, \ds \Big].\label{energy}
\end{align}
For $\bsigma_\Gamma= \bsigma_\Gamma^0$ the same relation holds, with $\lambda_\Gamma=\mu_\Gamma=0$. 
\end{lemma}
\begin{proof}
We choose $\mbf u$ as the test function in the first equation \eqref{e:vari1}. From the second equation we get $b(\bu,p)=0$. We have
\begin{align}
  \frac12  \frac{d}{dt}\int_\Omega \rho \|\bu\|^2 \, dx &= \frac12 \sum_{i=1}^2\frac{d}{dt}\int_{\Omega_i(t)} \rho \|\bu\|^2 \, dx  \nonumber \\ & = \frac12 \sum_{i=1}^2 \int_{\Omega_i(t)} \rho_i \frac{\partial}{\partial t}(\bu \cdot \bu) +\rho_i \bu \cdot \nabla (\bu \cdot \bu) \, \dx  \nonumber \\
 & = \sum_{i=1}^2 \int_{\Omega_i(t)} \rho_i \frac{\partial \bu}{\partial t}\cdot \bu + \rho_i (\bu \cdot \nabla \bu)\cdot \bu \, \dx = m(\partial_t \bu,\bu)+ c(\bu;\bu,\bu) \nonumber \\
& = -a(\bu,\bu) + f_\Gamma(\bu,\bu) + f_L(\bu).\label{h1}
\end{align}
We consider the three terms in \eqref{h1}. For the first one we get
\begin{equation} \label{h2}
 a(\bu,\bu)=\frac12 \int_\Omega \mu \, \tr \big( \bD (\bu)^2\big) \, \dx+ \int_{\partial \Omega_S} \beta_S \|\bPS \bu\|^2 \, \ds + \int_L \beta_L (\bu \cdot \bn_L)^2 \, \ds.
\end{equation}
We consider the second term $f_\Gamma(\bu,\bu)$ for $\bsigma_\Gamma= \bsigma_\Gamma^{BS}$. The case $\bsigma_\Gamma=\bsigma_\Gamma^0$ is a special case of this, with $\lambda_\Gamma= \mu_\Gamma=0$. We use $\divG \bu = \tr(\nablaG \bu)=\tr(\bPG \nablaG \bu)=\bPG : \nablaG \bu $ and $\tr \big((A+A^T)A\big)= \frac12 \tr\big((A+A^T)^2\big)$ for any matrix $A$. We obtain
\begin{equation}\label{h3}
 \begin{split}
   & f_\Gamma(\bu,\bu) = - \int_\Gamma \bsigma_\Gamma^{BS} : \nablaG \bu \, \ds \\ &  = - \tau \int_\Gamma \bPG : \nablaG \bu \, \ds - \lambda_\Gamma \int_\Gamma (\divG \bu)\bPG : \nablaG \bu \, \ds - \mu_\Gamma \int_\Gamma \bD_\Gamma(\bu) : \nablaG \bu \, \ds \\
& = - \tau \int_\Gamma \divG \bu \, \ds - \lambda_\Gamma \int_\Gamma \tr (\nablaG \bu)^2 \, \ds - \frac12 \mu_\Gamma \int_\Gamma \tr \big(\bD_\Gamma(\bu)^2 \big) \, \ds \\
 & = - \tau \frac{d}{dt} |\Gamma(t)| - \lambda_\Gamma \int_\Gamma \tr (\nablaG \bu)^2 \, \ds - \frac12 \mu_\Gamma \int_\Gamma \tr \big(\bD_\Gamma(\bu)^2 \big) \, \ds 
 \end{split}
\end{equation}
where in the last equality we used $\int_{\Gamma(t)} \divG \bu \, \ds = \frac{d}{dt} \int_{\Gamma(t)} 1 \, \ds$. For treating the third term $f_L(\bu)$ we use Stokes theorem in the plane that contains the wall $\partial \Omega_S$. The divergence operator in this plane is denoted by ${\rm div}_S$. Using $\Gamma_S^{(1)} \subset \partial \Omega_S$ and $L=\partial \Gamma_S^{(1)}$ we get
\begin{equation} \label{h4}
 \begin{split}
  f_L(\bu) &= \tau \cos \theta_e \int_L \bu \cdot \bn_L \, \ds  =\tau \cos \theta_e \int_{\Gamma_S^{(1)}} {\rm div}_S \bu \, \ds \\
 &= \tau \cos \theta_e \frac{d}{dt} \int_{\Gamma_S^{(1)}(t)} 1 \, \ds = \tau \cos \theta_e \frac{d}{dt} |\Gamma_S^{(1)}(t)| \\ & = - \frac{d}{dt} \big( \tau_1|\Gamma_S^{(1)}(t)|+ \tau_2|\Gamma_S^{(2)}(t)|\big) .
 \end{split}
\end{equation}
In the last inequality we used Young's relation $\tau \cos \theta_e = \tau_2 - \tau_1$  and $|\Gamma_S^{(1)}(t)|+ |\Gamma_S^{(2)}(t)|= |\partial \Omega_S|$ is independent of $t$. 
Combination of \eqref{h1},\eqref{h2},\eqref{h3} and \eqref{h4} completes the proof.
\end{proof}
\ \\[1ex]
A similar result, for the case $\bsigma_\Gamma=\bsigma_\Gamma^0$, is derived in \cite{Buscaglia2011, gerbeau2009}. The terms on the right hand-side in \eqref{energy} have obvious interpretations: they correspond to bulk viscosity, interface viscosity and wall and contact line dissipation energies. From $\mu >0, \, \lambda_\Gamma \geq 0,\, \mu_\Gamma \geq 0, \beta_S \geq 0, \, \beta_L \geq 0$ it follows that there is energy decay.  
\begin{remark} \rm
 The condition $\lambda_\Gamma \geq 0$ may  not be fulfilled in certain cases. A weaker  assumption that is more appropriate, cf.~\cite{Slattery} Sect. 4.9.5, is the following:
\begin{equation}  \label{gencond}
\lambda_\Gamma \geq - \mu_\Gamma.
 \end{equation}
With  $A:= \bPG \nablaG \bu   \bPG$ the surface viscosity part in \eqref{energy} can be written as 
\[
  \int_\Gamma \lambda_\Gamma \tr(A)^2 + \frac 12 \mu_\Gamma \tr \big((A+A^T)^2\big) \, \ds.
\]
Recall  that $(B,C) \to \tr (B^T C)$ defines a scalar product on $\mathbb{R}^{n\times n}$ and thus the Cauchy-Schwarz inequality $|\tr (B^T C)| \leq \tr(B^TB)^\frac12 \tr(C^T C)^\frac12$ holds. This yields
\[
 |\tr (A)| = \frac12 |\tr(A+A^T)| \leq \frac12 \tr\big((A+A^T)^2\big)^\frac12 \tr(\bPG)^\frac12 
\]
Using this in combination with $\tr(\bPG)=2$ and the assumption \eqref{gencond} we get
\[
  \lambda_\Gamma \tr(A)^2 + \frac 12 \mu_\Gamma \tr \big((A+A^T)^2 \geq (\lambda_\Gamma + \mu_\Gamma)\tr(A)^2 \geq 0
\]
and thus we conclude that the weaker  assumption  \eqref{gencond} suffices to have energy decay.
\end{remark}

%\section{Numerical method} \label{sectNumerical}
%In this section, we will firstly introduce the level set method for the interface tracking. The discretization of level set equation and interface reconstruction will be %explained briefly.
%To derive a space-time discrete equation of \eqref{e:vari1}, we will first apply a time discretization and then a space discretization, namely a Rothe approach.
%We will show XFEM method to handle the discontinuity for pressure and the Nitsche technique for weakly imposing the Navier condition.
%The coupling of Navier stokes equation and the level set equation will be explained at the end of this section.
%A Nitche technique is used to weakly impose $\mbf u\cdot \mbf n=0$ on the boundary $\partial\Omega_S$, which makes the implementations easier, can handle more general geometry and also help to avoid the Babuska paradox\cite{verfurth1986} for slip boundary conditions on curved domain.
%In DROPS, we use a level-set method to capture the movement of the interface as well as the contact line, and it also gives the information concerning the dynamic angle $\theta_d$.

\section{Level-set method} \label{sectLS}
%In DROPS, we use a level-set method to capture the movement of the interface as well as the contact line, and it also gives the information concerning the dynamic angle $\theta_d$.
A key difficulty in the numerical simulation of two-phase flows is the numerical approximation of the (implicitly given) interface. Different techniques are available, cf.~\cite{GrossReuskenBook} for an overview. In the numerical method used in this paper we use the popular level set technique, which implicitly captures the position of $\Gamma(t)$.  We will briefly introduce the method, and explain how one can  approximate the interface $\Gamma$. 
We introduce  a time-dependent level-set function $\phi(x,t)$ such that $\Gamma(t)=\{\,( x,t )~|~\phi(x,t)=0\,\}$. This level set function is given for $t=0$ (e.g. an approximate distance function to $\Gamma(0)$) and its evolution is determined by the linear transport equation
\begin{align}
\frac{\partial\phi}{\partial t}+\mbf u\cdot\nabla \phi=0.\lbl{e:levelset}
\end{align}
For the space discretization of this equation we use a standard finite element approach. Let $\{\cT_h\}_{h >0}$ be a shape regular family of tetrahedral triangulations of the domain $\Omega$. For given boundary data $\phi_D$ on the inflow boundary $\partial \Omega_{in}$ we introduce the  finite element space of piecewise quadratics:
$$V_h(\phi_D)=\{\, \phi_h\in C(\Omega)~|~ \phi_h|_T \in \mathcal P_2 ~ \text{for all}~T \in \cT_h,~~\phi_h=\phi_D~\text{on}~\partial \Omega_{in}\,\}.$$
The level set equation is semi-discretized using the streamline diffusion finite element method (SDFEM): determine $\phi_h(\cdot,t) \in V_h(\phi_D)$ such that 
\begin{align}\lbl{e:lsetSUPG}
\sum_{T \in \cT_h} \big(\frac{\partial\phi_h}{\partial t}+\mbf u \cdot\nabla\phi_h,v_h+\delta_T\mbf u\cdot\nabla v_h\big)_{L^2(T)}=0 \quad \text{for all}~v_h \in V_h(0),
\end{align}
where  $\delta_T$ is the SDFEM stabilization parameter, cf.~\cite{TobiskaBook}. This parameter typically has the form $\delta_T = c \frac{h_T}{\max \{ \epsilon_0, \|\bu\|_{\infty,T} \} }$ with $\epsilon_0 >0$ and $c = \mathcal{O}(1)$.
%where the mesh Pelect parameter $\delta(x)=c h_K/\max(\epsilon_0,\|u_h\|_{\infty,K})$ whenever $x\in K\in\mathcal{T}_h$,
%with a positive constant $c$ and a small tolerance number $\epsilon_0>0$.
The fact that we use quadratic (instead of linear) finite elements is essential for an accurate approximation of surface tension forces \cite{Gross06}. The heuristic explanation for this is that in the surface tension force the curvature of $\Gamma(t)$ plays a key role and approximation of the curvature using piecewise linears for $\phi_h(\cdot,t)$ turns out to be too inaccurate. In the discretization of the surface and line forces $f_\Gamma$ and $f_L$ in \eqref{e:vari1}, which is explained in section~\ref{sectforces} below, we need an explicitly accessible approximation of $\Gamma(t)$ and an approximation of the normal $n_\Gamma$. A crucial point in our approach is that the former is based on a piecewise \emph{linear} interpolation of the piecewise quadratic function $\phi_h(\cdot,t)$, whereas for the latter we use the piecewise \emph{quadratic} function. We give a more precise explanation of this approach. Computing (a parametrization of) the zero level of the piecewise quadratic function $\phi_h(\cdot,t)$ is computationally 
extremely expensive. Instead,  we  determine the piecewise linear interpolation of  $\phi_h(\cdot,t)$ on the regularly refined triangulation $\cT_{\frac12 h}$, denoted by $I\phi_h(\cdot,t)$. The approximate interface is given by
\begin{equation} \label{defGh}
 \Gamma_h(t):=\{\, x \in \Omega~|~I\phi_h(x,t)=0~\,\}. 
\end{equation}
This approximate interface is piecewise planar (consisting of triangles an quadrilaterals) and easy to determine. Under reasonable assumptions the estimate ${\rm dist}(\Gamma,\Gamma_h) \leq c h^2$ holds, cf.~\cite{GrossReuskenBook}. The normals and corresponding projections needed in the numerical approximation of the force terms, cf. section~\ref{sectforces}, are determined as follows (we delete the dependence on $t$ in the notation): 
\begin{equation} \label{defntilde}
\tilde{\mbf n}_h:=\frac{\nabla\phi_h}{\|\nabla\phi_h\|}, \quad \tilde{\mathbf{P}}_h=\mbf I-\tilde{\mbf n}_h\tilde{\mbf n}_h^T.
\end{equation}
This choice leads to significantly better results  than using the normals to the piecewise planar approximation $\Gamma_h$. \\
Related to the numerical treatment of the level set function there are further important issues such as  reinitialization (also called reparametrization) and volume correction. 
We do not treat these topics in this paper. For reinitialization we use a Fast Marching Method as presented  in \cite{Sethian96}. For volume correction  we apply a simple shift technique. For more information about these methods we refer to \cite{GrossReuskenBook}.

%\subsection{Time discretization}
%For the discretization of \eqref{e:vari1}, we will first look at the time discretization.  
%We introduce the notation
%\begin{equation}
% r(\bu,\bv, t) :=f_{ext}(\mbf v)+f_{\Gamma}(\bu,\mbf v) +f_{L}(\mbf v)-a(\mbf u, \mbf v)-c(\mbf u;\mbf u,\mbf v).
%\end{equation}
%The variational formulation \eqref{e:vari1} is reformulated as: find $\bu(t) \in \mbf X_t, p \in Q$ such that 
% \begin{equation}
% \left\{
% \begin{array}{ll}
% m(\partial_t\mbf u,\mbf v) + b(\mbf v, p)=  r(\bu,\bv, t)
%  \\
%b(\mbf u, q) =0
% \end{array}
%\right.\lbl{e:vari2}
%\end{equation}
%for all $\bv \in \mbf X_t$, $q \in Q$. 
%We will introduce a semi-implicit Euler method in the sense that the $\bu$ part is treated implicit but the other time dependent parts are treated explicitly.  
%To be more precise, we consider the following problem: Given $\bu_0 \in \mbf X_0, \text{for } n \geq 0, \text{determine } \bu^{n+1} \in \mbf X_{t^{n}} , p \in Q$ such that
% \begin{equation}
% \left\{
% \begin{array}{ll}
% \int_\Omega \rho(t^n) \frac{\bu^{n+1} - \bu^{n}}{\Delta t} \bv dx  + b(\mbf v, p)=  r(\bu^{n+1},\bv, t^n)
%  \\
%b(\mbf u^{n+1}, q) =0
% \end{array}
%\right.\lbl{e:implEuler}
%\end{equation}
%for all $\bv \in \mbf X_{t^{n}}$, $q \in Q$.

\section{Pressure discretization: stabilized $P_1$-XFEM method} \label{sectXFEM}
\begin{figure}[ht!]
\begin{center}
\begin{overpic}[scale=0.4, angle=90]{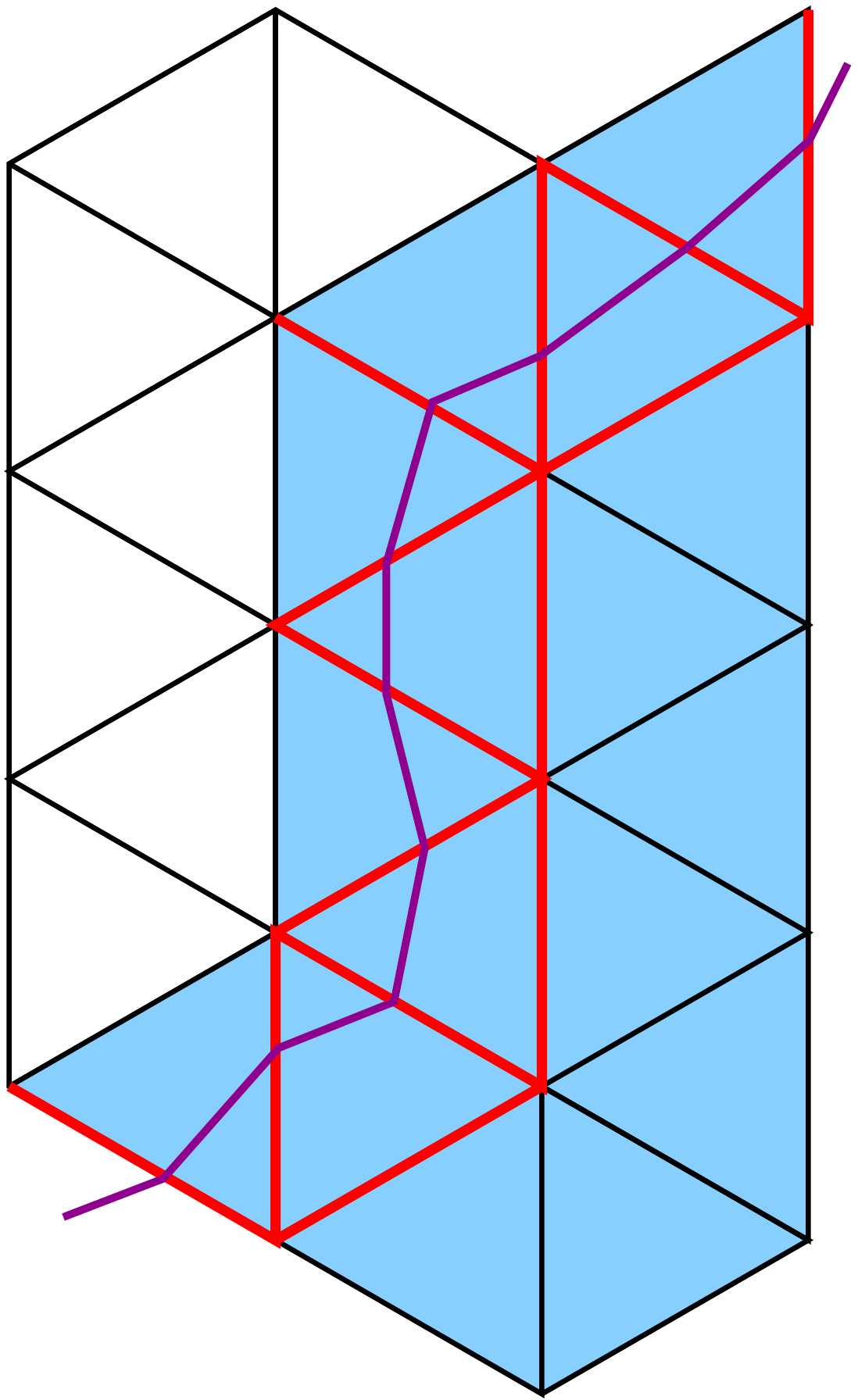}
   \put(16,59){$\mathcal{F}_1$}
  \put(87,6){$\Gamma_h$}
  \put(40,32){$\Omega_{1,h}$}
\end{overpic}
\caption{The subdomain $\Omega_{1,h}$ (in blue), the set of faces $\mathcal{F}_1$(in red) and the reconstructed interface $\Gamma_h$ (in purple).}
\end{center}
\label{fig:frictiousDomain}
\end{figure}

In the two-phase problem the normal bulk stress tensor $\bsigma \bn_\Gamma$ is discontinuous across the interface $\Gamma$, cf.~\eqref{e:Gamma}. 
This results in a pressure jump   and a discontinuity in the velocity derivative across the interface. 
An accurate approximation of these discontinuities is a difficult task because, due to the use of the level set technique, the triangulation is not fitted to the (moving) interface. In particular the discontinuity in the pressure results in very large discretization errors (of order $\sqrt{h}$ in the $L^2$-norm) if standard finite element spaces are used. 
In recent years  the XFEM technique has been successfully  applied to deal with this problem. For an accurate pressure discretization
we will use a $P_1$-XFEM known  from the literature \cite{Hansbo14}, which we now outline . We use the same family of shape-regular tetrahedral triangulations $\{\cT_h\}_{h>0}$ as for the discretization of the level set equation.
The space of piecewise linears is denoted by
\begin{align*}
Q_h&:=\{q\in C({\Omega}): q|_T\in \mathcal P_1 ~\text{for all}~T\in\mathcal{T}_h \}.%\hbox{ if } area(\Gamma_N)=0
\end{align*}
%It is well-known that the Hood-Taylor finite element space is stable for (Navier-) Stokes problems in the sense that it satisfies
%the well-known Babuska-Brezzi inf-sup condition. The convergence order is also optimal for one phase flow. The $L^2$ errors for velocity decay with order $O(h^3)$ and that for pressure decay with $O(h^2)$. However, the convergence behavior is much worse for two-phase flows because of the discontinuity of pressure  and the non-smoothness of the velocity across the interface $\Gamma(t)$. 
%However as we mentioned before, we will use the standard space $X_h$ as our discretized space for velocity.  
%To improve the accuracy of the pressure approximation, an extended finite element method introduced in \cite{Gross06b} is used for pressure. 
%One drawback of XFEM is that it is not LBB-stable, so certain stabilization techniques are required to apply this method.
%One problem of this method is that you need to find an appropriate criterion to define "small`` in the sense it gives satisfactory results in the meshes you use. 
%To our knowledge this XFEM technique has not been applied for the simulation of two-phase flows with dynamic contact lines, yet. In the following we briefly explain the idea of this technique.
%To avoid technical details, we make a generic intersection assumption that $\Gamma \cap T$ does not coincide with an edge of $T$. 
We introduce the subdomains
\begin{equation*}
 \Omega_{i,h} := \{ T \in \mathcal{T}_h : \text{meas}_3(T\cap\Omega_i)>0\}, \qquad i =1, 2,
\end{equation*}
and the corresponding  finite element spaces
\begin{equation*}
 Q_{i,h} := {Q_h}_{|\Omega_{i,h}}=\{\, q\in C(\Omega_{i,h})~|~q|_T\in \mathcal P_1~ \text{for all}~ T\in \Omega_{i,h}\, \}, \qquad i=1,2.
\end{equation*}
%We use the same notation$\Omega_{i,h}$ for the set of simplices as well as for the subdomain of $\Omega$ which is formed by these simplices.
We define $\mathcal{T}^\Gamma_h := \{\, T \in \mathcal{T}_h~|~ \text{meas}_2 (T \cap \Gamma_h) > 0 \, \}$, which is the set of elements intersected by $\Gamma_h$.
The following  sets of faces are needed in the stabilization procedure, cf.~Fig.~\ref{fig:frictiousDomain}:
\begin{equation*}
 \mathcal F_i = \{\, F \subset \partial T~|~T \in \mathcal{T}^{\Gamma}_h, F \not \subset \partial \Omega_{i,h} \,\},~~ i = 1,2,
\end{equation*}
and $\mathcal{F}_h := \mathcal{F}_1 \cup \mathcal{F}_2$. For each $F \in \mathcal{F}_h$, the unit normal  with a fixed orientation is denoted by $n_F$. 

A given pair $p_h=(p_{1,h}, p_{2,h}) \in Q_{1,h} \times Q_{2,h}$ may have
 two values for $x \in \mathcal{T}^{\Gamma}_h$. We define a uni-valued function $p^\Gamma_h(x) \in C(\Omega_1 \cup \Omega_2)$ by 
\begin{equation*}
 p^{\Gamma}_h (x) = p_{i,h}(x), \qquad \text{for } x \in \Omega_i.
\end{equation*}
The mapping $p_h \longmapsto p^\Gamma_h$ is bijective. 
The $P_1$-XFEM space  is defined by 
\begin{equation}
 Q^\Gamma_h := (Q_{1,h} \times Q_{2,h})/ \mathbb{R}= \{\, p_h \in Q_{1,h} \times Q_{2,h}~|~ \int_\Omega p_h^\Gamma\, dx=0\, \}.
\label{xfemspace}
\end{equation}
Note that $\{p^\Gamma_h : p_h \in Q^\Gamma_h \}$ is a subspace of the pressure space $Q$. The space $Q_h^\Gamma$ can also be characterized as a space obtained by adding certain discontinuous basis functions, with support in $\cT_h^\Gamma$, to the original space $Q_h$. This explains the name ``extended FEM''. Due to the very small areas of certain cut elements $T \cap \Omega_i$ a discretization in the space $Q^\Gamma_h$ may lead to instabilities in the discrete solution and/or very poor conditioning of the stiffness matrix, cf.~\cite{Hansbo14,Reusken07}. Recently, the following ghost-penalty stabilization for $P_1$-XFEM has been introduced \cite{Hansbo14}:
\begin{align}
\begin{split}
 j(p_h, q_h) &=\sum^2_{i=1} j_i(p_{i,h}, q_{i,h}), \qquad p_h, q_h \in Q_{1,h} \times Q_{2,h}, \\
 j_i(p_{i,h}, q_{i,h}) &= \mu_i^{-1} \sum_{F \in \mathcal{F}_i} h^3_F ([\nabla p_{i,h} \cdot n_F][ \nabla q_{i,h} \cdot n_F])_{0, F},
\label{ghostpenalty}
\end{split}
\end{align}
where $h_F$ is the diameter of the face $F$ and $[\nabla p_{i,h} \cdot n_F]$ denotes the jump of the normal components of the piecewise constant function $\nabla p_{i,h}$ across the face $F$. Below in section~\ref{sectFull} we will use the space $Q^\Gamma_h$ for a conforming pressure discretization and modify the bilinear form by adding the stabilization term $\epsilon_p j(\cdot,\cdot)$ with the parameter $\epsilon_p>0$. Note that this is a consistent stabilization, in the sense that for a piecewise continuous pressure solution $p$ (i.e., $p_{|\Omega_i}$ is continuous)  and for all $q_h \in Q^\Gamma_h$ we have $j(p,q_h)=0$.

%The term \eqref{ghostpenalty} is also referred to as a ghost penalty term. Now we will continue with the Nitsche technique and come back to the XFEM method later in the space-time discrete equations.
Another  XFEM   stabilization approach is presented in \cite{Reusken07}. In that method one deletes from the XFEM space all  basis functions with a relatively very small support. In our experience the ghost penalty method performs better. Hence, in the numerical experiments we use the stabilization method based on \eqref{ghostpenalty}. 
%The traditional way to introduce exteneded finite element
%\begin{comment}
%Define the index set $\mathcal{J} =\{ 1, 2, \ldots, n\}$, where $n= \text{dim}Q_h$. Let $(q_j)_{j \in \mathcal{J}}$ be the nodal basis of $Q_h$.
%We define $\mathcal{J}_\Gamma := \{j \in \mathcal{J} : \text{meas}_2(\Gamma \cap \supp q_j)>0\}$ as the index set of the basis function whose support is intersected with $\Gamma$.
%The Heaviside function $H_\Gamma$ has the values $H_\Gamma(x)=0$ for $x\in \Omega_1$ and $H_\Gamma(x)=1$ for $x\in \Omega_2$, which is a characteristic function of the domain %$\Omega_i, i=1, 2$.
%We define further a so-called enrichment function $\Phi_j(x) := H_\Gamma(x) -  H_\Gamma(x_j), j\in \mathcal{J}_\Gamma$.
%The additional basis function can be constructed as $q^\Gamma_j(x)= q_j(x)\Phi_j(x), j\in \mathcal{J}_\Gamma$.
%These functions are discontinuous across $\Gamma$.
%The extended finite element space is written as
%\begin{align*}
%Q_h^X&=Q_h \oplus \text{span}\{q^\Gamma_j: j \in \mathcal{J}_\Gamma\}.
%\end{align*}
%More details is referred to \cite{Gross06b,GrossReuskenBook}.
%\end{comment}
%In our computations, the finite element pair $(X_h,Q_h^X)$ is utilized. However, we would like to remark that our approach also works for other stable finite element pairs.

\section{Velocity discretization: Nitsche $P_2$-FEM} \label{sect|Nitsche}
As noted above, the discontinuity of $ \bsigma \bn_\Gamma$ across the interface causes a discontinuity not only in the pressure but also in the normal derivative of the velocity, if $\mu_1 \neq \mu_2$. Hence, if $\mu_1 \neq \mu_2$ a standard $P_2$ finite element space for velocity discretization does not have optimal approximation properties. Numerical experiments \cite{Sauerland2011,GrossReuskenBook} have shown that in cases  with small viscosity jumps often the velocity error in the $P_2$-FEM space does not dominate the total discretization error. This explains why in practice  the  pair $P_1$-XFEM (with stabilization) for pressure and $P_2$-FEM for velocity is often used. In this paper we also use this pair, i.e.,  velocity is approximated in the  finite element space
\begin{align} \label{velspace}
X_h&:=\{\, \mbf v\in C({\Omega})~|~ \mbf v|_T\in (\mathcal P_2)^3~~\text{for all}~ T\in\mathcal{T}_h, ~\bv_{|\partial \Omega_D} =0 \,\}.
\end{align}
Note that this is \emph{not} a conforming space, because the Navier boundary condition $\bn_S \cdot \mbf v =0$ on $\partial \Omega_S$ is not an essential boundary condition in the finite element space. 
Often the direct imposition of this condition is not easy to implement, especially when the normal $\bn_S$ does not have the direction of a  coordinate axis (e.g., an inclined wall or a curved boundary $\partial \Omega_S$). In  this paper we show how one can use  
the Nitsche technique   to weakly impose $\mbf u\cdot \mbf n_S=0$ on the boundary $\partial\Omega_S$  \cite{Hansbo05a,Nitsche71}.
%In addition, Babuska's paradox for slip condition on curved boundary
%might make the method fails\cite{verfuerth1987}. (We will not  discuss further the topic since the domain $\Omega$ is assumed polygonal in this paper.
%Further studies will be illustrated in an alternative paper.) To overcome these drawbacks, we consider using the Nitche-technique,

For this Nitsche method  we have to modify the blinear form(s) used in the variational formulation. We recall the arguments used in section~\ref{sectvariational}.   
Note that the test functions $\bv \in \mbf X_h $ do not satisfy $\bv \cdot \bn_S=0$.
Following the arguments in section~\ref{sectvariational} we observe that \eqref{force1} is still correct, but now the term $\int_{\partial \Omega_S} (\bn_S \cdot \bsigma \bn_S) (\bv \cdot \bn_S) \, \ds$ does not vanish. This term can be written as
\begin{equation*} \label{addi1}
 \int_{\partial \Omega_S} (\bn_S \cdot \bsigma \bn_S) (\bv \cdot \bn_S) \, \ds = -\int_{\partial \Omega_S} p \,\bv \cdot \bn_S \, \ds + \int_{\partial \Omega_S} \mu (\bn_S \cdot \bD(\bu) \bn_S) (\bv \cdot \bn_S) \, \ds.  
\end{equation*}
The first term on the right hand-side will be added to the bilinear form $b(\cdot,\cdot)$ and the second term will be added to $a(\cdot,\cdot)$. Similarly following the arguments in the second step of integration by parts (starting from \eqref{partial2}) we observe that \eqref{ref} is still correct, but now the term $\int_L (\bn_S \cdot \bsigma_\Gamma \btau_L)(\bv \cdot \bn_S)\, \ds $ does not vanish. 
This term will be added to the contact line functional $f_L$. 

Hence, in  the Nitsche method  we introduce three modified bilinear forms:
\begin{align}
\tilde b(\mbf v, q) &= b(\mbf v, q)+ \int_{\partial\Omega_S}(\mbf v\cdot \mbf n_{S}) q\, \ds \label{modb},\\
\tilde{a}(\mbf u,\mbf v)&:=a(\bu,\bv) -\int_{\partial \Omega_S}\mu(\mbf n_S \cdot\mbf D(\mbf u)\mbf n_S)(\mbf v\cdot \mbf n_S)\, \ds \nonumber \\ & ~~ -\int_{\partial \Omega_S}\mu(\mbf n_S \cdot\mbf D(\mbf v)\mbf n_S)(\mbf u\cdot \mbf n_S)\, \ds 
 + \frac{\alpha}{h} \int_{\partial \Omega_S} (\mbf u\cdot \mbf n_S)(\mbf v\cdot\mbf n_S)\,\ds, \label{moda}\\
 \tilde f_{L}(\bu, \mbf v) &= f_L(\bv)+\int_L (\bn_S \cdot \bsigma_\Gamma \btau_L)(\bv \cdot \bn_S)\, \ds .\label{modfL}
\end{align}
In the bilinear form $\tilde a(\cdot,\cdot)$, besides the correction term $\int_{\partial \Omega_S} \mu (\bn_S \cdot \bD(\bu) \bn_S) (\bv \cdot \bn_S) \, \ds$, there are two more terms in \eqref{moda}. The first one, $\int_{\partial \Omega_S}\mu(\mbf n_S \cdot\mbf D(\mbf v)\mbf n_S)(\mbf u\cdot \mbf n_S)\, \ds$, is added to maintain symmetry.
The second term $\frac{\alpha}{h} \int_{\partial \Omega_S} (\mbf u\cdot \mbf n_S)(\mbf v\cdot\mbf n_S)\,\ds$ is a penalty term that enforces the no-penetration condition $\bu \cdot \bn_S=0$ in a weak sense. The parameter $\alpha >0$ is independent of $h$ and has to be taken sufficiently large. This penalty term is the usual one in Nitsche's method and it is consistent in the sense that it vanishes if $\bu$ is the solution of the continuous problem (which satisfies $\bu \cdot \bn_S=0$). Due to this we have that the error $(\bu - \bu_h,p-p_h)$ has the usual Galerkin orthogonality property w.r.t. $X_{h}\times Q_h^\Gamma$.

Using  $\btau_L= \frac{\bPG \bn_S}{\|\bPG \bn_S\|}$ and $\bn_L= \frac{\bPS \bn_\Gamma}{\|\bPS \bn_\Gamma\|}$ the right-hand side  in  \eqref{modfL} can be rewritten and we obtain 
\begin{equation} \label{rewrt}
\tilde f_{L}(\bu, \mbf v)=\cos \theta_e \int_L \tau\|\bPS \bn_\Gamma\|^{-1} \bv \cdot \bPS \bn_\Gamma\, \ds +\int_L  \|\bPG \bn_S\|^{-1} (\bn_S \cdot \bsigma_\Gamma \bn_S)(\bv \cdot \bn_S)\, \ds.
\end{equation}
In the latter representation the (only) geometric information we need is the location of the contact line $L$, the known normal $\bn_S$ and the surface normal $\bn_\Gamma$.
\begin{remark} \rm 
%In \eqref{modfL} the functional depends on $\bu$ if $\bsigma_\Gamma$ depends on $\bu$ (e.g., $\bsigma_\Gamma=\bsigma_\Gamma^{BS}$). 
For the case $\bsigma_\Gamma=\bsigma_\Gamma^0=\tau \bPG$ the formula in \eqref{rewrt} simplifies to 
\begin{equation}  \label{specfl} 
 \tilde f_{L}(\bu, \mbf v) =\tilde f_{L}^0(\mbf v) : = \cos \theta_e \int_L \tau\|\bPS \bn_\Gamma\|^{-1} \bv \cdot \bPS \bn_\Gamma\, \ds +\int_L \tau \|\bPG \bn_S\|  \bv \cdot \bn_S \, \ds .
\end{equation}
\end{remark}

%determine $(\mbf u^{n+1}_h,p^{n+1}_h)\in \mbf X_{h}\times Q_h^\Gamma$ such that
%\begin{equation} \label{Nitsche} \begin{split}
%  & m(\frac{\mbf u_h^{n+1}-\mbf u_h^{n}}{\Delta t},\mbf v_h)+\tilde{a}(\mbf u^{n+1}_h,\mbf v_h)+\tilde {b}(\mbf v_h, p^{\Gamma, n+1}_h)+c(\mbf u^{n+1}_h;\mbf u^{n+1}_h,\mbf %v_h) + j(p^{n+1}_h, q_h)
%   \\&=f_{ext}(\mbf v_h)+\tilde f_{\Gamma}(\bu^{n+1}_h,\mbf v_h)  + \tilde f_{L}(\bu^{n+1}_h,\mbf v_h)
%       \quad \text{for all}~~ \mbf v_h\in \mbf X_h, \\
%&  \tilde b(\mbf u^{n+1}_h, q^{\Gamma}_{h}) =0,\qquad  \text{for all}~~ q_h\in Q^\Gamma_h.
%\end{split}
%\end{equation}

%\begin{remark} {\bf AR: can this be changed: use stabilized XFEM always?}\\
% In our experiments, we only applied the stabilized XFEM method for stationary problem, because the preconditioning of the non-stationary problem is not implemented at that time. In the following 
% formula, we will take out the term  $j(p_h, q_h)$. For the stabilization of the non-stationary problem, we take the traditional reduced XFEM method introduced in \cite{GrossReuskenBook}. 
%\end{remark}

\section{Discretization of force terms} \label{sectforces}
The interface and contact line forces are given by:
\begin{align}
 f_{\Gamma}(\bu, \mbf v)&=-\int_{\Gamma} \bsigma_\Gamma:\nabla_{\Gamma} \mbf v\, \ds, \label{f1} \\
\tilde f_{L}(\bu, \mbf v)& =\cos \theta_e \int_L \tau\|\bPS \bn_\Gamma\|^{-1} \bv \cdot \bPS \bn_\Gamma\, \ds +\int_L  \|\bPG \bn_S\|^{-1} (\bn_S \cdot \bsigma_\Gamma \bn_S)(\bv \cdot \bn_S)\, \ds. \label{f2}
\end{align}
We discuss the discretization of these (bi)linear forms. The geometry $\Gamma=\Gamma(t)$ and $L=L(t)$ are approximated as explained in section~\ref{sectLS}. For the interface we use the approximation $\Gamma_h$ given in \eqref{defGh}. For the contact line approximation we take $L_h:=\Gamma_h \cap \partial\Omega_S$. Besides the known normal $\bn_S$, the only further geometric information we need in  \eqref{f1}-\eqref{f2} is the surface normal $\bn_\Gamma$. This normal is approximated by $\tilde{\mbf n}_h$, with corresponding projection denoted by $\tilde{\mbf P}_h$,  as in \eqref{defntilde}.

For the case $\bsigma_\Gamma=\bsigma_\Gamma^0=\tau \bPG$, cf.~\eqref{specfl}, we obtain the following discrete functionals, with  $\mbf e_i$ the $i$th basis vector in $\mathbb{R}^3$:
\begin{align}
f_{\Gamma,h}^0(\mbf v_h)& =-\sum_{i=1}^3\int_{\Gamma_h} \tau \tilde{\mbf P}_h \mbf e_i\cdot\nabla (\mbf v_h)_i\ds \label{tt1} \\
\tilde f_{L,h}^0(\mbf v_h) &  = \cos \theta_e \int_{L_h} \tau\|\bPS \tilde \bn_h\|^{-1} \bv_h \cdot \bPS \tilde \bn_h\, \ds +\int_{L_h} \tau \|\tilde{\mbf P}_h  \bn_S\|  \bv_h \cdot \bn_S \, \ds . \label{tt2}
\end{align}
For the case $\bsigma_\Gamma=\bsigma_\Gamma^{BS}$ similar results are easily obtained. The resulting discrete bilinear forms are denoted by $f_{\Gamma,h}(\mbf u_h,\mbf v_h)$, $ \tilde f_{L,h}(\mbf u_h, \mbf v_h)$. 
\section{Fully discrete problem} \label{sectFull} 
We apply a Rothe approach in which we first apply an implicit Euler method to the fully coupled system of level set and Navier-Stokes equations and then use the methods explained in the previous sections for a spatial discretization. Thus we obtain the following fully discrete problem, per time step. To simplify the presentation, we assume $\delta_T= \delta_h$ for all $T \in \cT_h$ in the SDFEM. Determine $(\mbf u_h^{n+1},p_h^{n+1},\phi_h^{n+1})\in \mbf X_h\times Q_h^\Gamma\times V_h(\phi_D)$ such that
\begin{align}
 & m(\frac{\mbf u_h^{n+1}-\mbf u_h^{n}}{\Delta t},\mbf v_h)+\tilde{a}(\mbf u^{n+1}_h,\mbf v_h)+c(\mbf u^{n+1}_h;\mbf u^{n+1}_h,\mbf v_h)+\tilde{b}(\mbf v_h, p^{n+1, \Gamma}_h) \nonumber \\
 & + \tilde b(\mbf u^{n+1}_h, q^{\Gamma}_h) + \epsilon_p j( p_h^{n+1}, q_h)=f_{ext}(\mbf v_h)+ f_{\Gamma,h}(\bu^{n+1}_h, \mbf v_h)  +\tilde f_{L,h}(\bu^{n+1}_h, \mbf v_h)  \lbl{e:decoupled1} \\
    & \qquad \qquad \qquad\qquad \qquad \qquad \qquad \text{for all}~\mbf v_h\in\mbf X_h,~q_h\in Q^\Gamma_h, \nonumber \\
&(\frac{\phi_h^{n+1}-\phi_h^{n}}{\Delta t}+\mbf u^{n+1}_h\cdot\nabla\phi^{n+1}_h ,\psi_h+\delta_h \mbf u^{n+1}_h\cdot\nabla\psi_h)_{L^2(\Omega)}=0, \quad\forall \psi_h\in V_h(0).\lbl{e:decoupled3}
\end{align}
For the case $\bsigma_\Gamma=\bsigma_\Gamma^0=\tau \bPG$ the discrete force terms $ f_{\Gamma,h}$ and $\tilde f_{L,h}$ depend only on $\mbf v_h$ and are given in \eqref{tt1}, \eqref{tt2}. We emphasize that through the dependence of $f_{\Gamma,h}$  and $\tilde f_{L,h}$ on $\Gamma_h=\Gamma_h(t)$, which is the zero level of $I\phi_h(\cdot,t)$, there is a strong nonlinear coupling between the equations in  \eqref{e:decoupled1} (fluid dynamics) and in \eqref{e:decoupled3} (interface and contact line dynamics). The decoupling of these two equations is a delicate issue. 
In the following, we describe two decoupling approaches we used in our simulations. 
In the first approach(weak coupling), we decouple the computation of \eqref{e:decoupled1} and \eqref{e:decoupled3} in each time step. In the second approach(strong coupling), the two equations are solved coupled together by a fixed point iterative method in each step.
\subsubsection*{Weak coupling}
In each time step, we solve  \eqref{e:decoupled1} and \eqref{e:decoupled3} to get $(\mbf u_h^{n+1},p_h^{n+1})$ and $\phi^{n+1}_h$ based on $\phi^n_h$ and $\mbf u_h^n$ from the  previous time step.  In each time step the two subproblems, Navier-Stokes for fluid dynamics and the level set equation for the interface dynamics, are   decoupled. This approach is very simple and has relatively low computational costs per time step. In our experiments this weak coupling performs well in problems where the flow evolves to a stationary one and one is (mainly) interested in the stationary limit solution. In cases where the solution has a strong dependence on time, it is better to use the following strong coupling approach. 
\subsubsection*{Strong coupling}
In the strong coupling algorithm, per time step  we solve (with a certain tolerance) the coupled system \eqref{e:decoupled1}-\eqref{e:decoupled3} for the unknowns $(\mbf u_h^{n+1},p_h^{n+1},\phi_h^{n+1})$. This is realized by a certain fixed point approach which iterates between the equations in \eqref{e:decoupled1} and \eqref{e:decoupled3}. In this approach there is a special semi-implict treatment of the surface tension force term $f_{\Gamma,h}(\bu^{n+1}_h, \mbf v_h)$, with $\Gamma=\Gamma(\phi_h^{n+1})$, which is very important for the rate of convergence of this fixed point iteration. We do not explain this further here, but refer to \cite{GrossReuskenBook} for more information on the fixed point iteration.

After $ (\mbf u_h^{n+1},p_h^{n+1},\phi_h^{n+1})$ has been determined, it is 
checked whether a reinitialization or a  volume correction of the level-set function $\phi_h^{n+1}$ should be determined. Again, because it is outside the scope of this paper, we do not explain these methods, but refer to \cite{GrossReuskenBook}.

\section{Numerical experiments} \label{sectExp}
In this section we apply the solver, outlined in the sections above, to a few test problems. 
We restrict to  the interface stress tensor $\bsigma_\Gamma = \bsigma^0_\Gamma$, and for the contact line force $\mbf f_L$ we restrict to  $\beta_L =0$, i.e., $\mbf f_L=\tau \cos \theta_e \bn_L$, cf.~\eqref{FL}.
This  GNBC model, cf. Remark~\ref{remmodels}, may have   limitations in reproducing certain (macroscopic) physical observations \cite{Buscaglia2011}. However, as mentioned above, in this paper we do not study the quality of different models for describing contact line dynamics.
  
We summarize the main components of the solver:
\begin{itemize}
 \item We use a modified $P_2$-$P_1$ Hood-Taylor pair for the spatial  discretization of the Navier-Stokes equation. The modification is that instead of $P_1$ for the pressure we use the stabilized $P_1$-XFEM discretization. We use a hierarchy of tetrahedral grids, which are locally refined close to the (moving) interface. 
\item We use the level set method for interface capturing. For the spatial discretization of the level set equation we use a standard streamline diffusion FEM (SDFEM) with $P_2$ finite elements.
  In this level set method it is essential to use a suitable reparametrization technique (which is not explained in this paper).
\item For the evaluation of surface tension and contact line forces we use a piecewise linear reconstruction of the interface. These forces are discretized as in \eqref{tt1}, \eqref{tt2}.
\item We use Nitsche's method for handling the Navier boundary condition, resulting in a modified bilinear form as in \eqref{moda}.
\item For the numerical time integration we use an implicit Euler method applied to the coupled system. Per time step either a weak or a strong coupling is used. If a strong coupling is used, one needs a suitable iterative solver for the coupled nonlinear system of equation (which is  not explained in this paper). For solving the resulting large linear sparse systems of equations tailor-made preconditioned Krylov subspace methods are used (not explained in this paper). 
\end{itemize}
In the test problems considered in section~\ref{sectdroplet} we use the weak coupling in the time integration, whereas in all other test problems we use the strong coupling. 
\subsection{Static and sliding droplets} \label{sectdroplet}
%In this subsection, we will first test the accuracy of our numerical method by considering a static droplet on a planar surface. 
%It is well-known that numerical methods will introduce spurious velocity for a stationary droplet problem\cite{Ganesan05} so that the discrete velocity is not exactly zero. One can use this fact to test %the convergence of a numerical method. 
%Our experiments shows the convergence order of our method is the same as that for a droplet without contact line. 
%In addition, we also compute a droplet which arrives at a stationary state from  an inequilibrium state. The accuracy in this case is satisfactory.
\subsubsection*{Static droplet on a plate} 
\begin{figure}[htb]
\begin{center}
  \begin{tikzpicture}[scale=1.9, font=\footnotesize]
    %\hspace*{0.8cm}
    \draw (3,0) -- (0,0) -- (0,2) -- (3,2)--cycle;
    \draw [decoration={border,segment length=0.9mm,amplitude=0.8mm,mirror,angle=45},decorate](0,0) -- (3,0);
    \filldraw [orange,thick,domain=30:150,opacity=0.5] plot ({1.5 + cos(\x)}, {-0.5+sin(\x)});
    \node at (1.5,0.3) {$\Omega 1$};
    \draw [domain=30:150, thick] plot ({1.5 + cos(\x)}, {-0.5+sin(\x)});
    \node at (1.5,1.0) {$\Omega 2$};
    \draw [domain=0:60] plot ({1.5 - 1*sin(60) + 0.1*cos(\x)}, {0.1*sin(\x)});
    \node at (1.0,0.095) [font=\tiny]{$\theta_0=60^{\circ}$};
    \draw [<->, thick] (1.5, {-cos(60)}) -- ({1.5 + cos(150)}, {-cos(60)  + sin(150)});
    \draw [<->, thick] (1.5, {-cos(60)}) -- ({1.5 + cos(150)}, {-cos(60)} );
    \draw [dashed] (1.5, 0) -- (1.5, {- cos(60)});
    \draw [dashed] ({1.5 + cos(150)}, 0) -- ({1.5 + cos(150)}, {-cos(60)});
    \node at (1.2, {-cos(60) + 0.27}){$R$};
    \node at (1, {-cos(60) - 0.10}){$r$};
  \end{tikzpicture}
  \caption{A sketch of the static droplet  system.}
\label{fig:9.1Setup}
\end{center}
\end{figure}
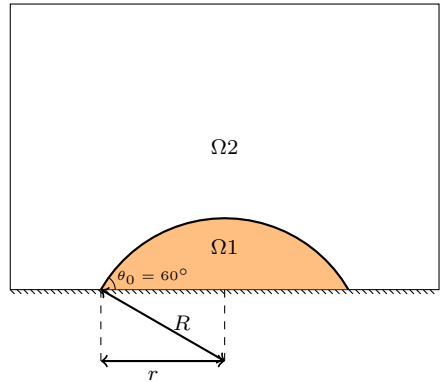

In two-phase incompressible flow simulations without contact line, a system with a static droplet is often used as a first test for the accuracy of the two-phase flow solver. In this basic test problem the quality of the discretization of surface tension forces and of the discontinuous pressure solution can be verified. We consider such a static droplet problem \emph{with} contact line.  
The test problem, sketched in Figure~\ref{fig:9.1Setup}, has a static droplet with radius $ R = 0.1$  on a planar surface in a cubic domain $\Omega = (0, 0.5)^3$. 
We assume a constant surface tension coefficient $\tau = 0.5$.
The densities and viscosities of both fluids are set to 1, and we set gravity to 0. 
On the whole boundary we assume a Navier slip condition with $\beta_S= 0.05$.

\begin{figure}[ht!]
\begin{center}
  \includegraphics[width=0.8\textwidth,height=0.35\textheight]{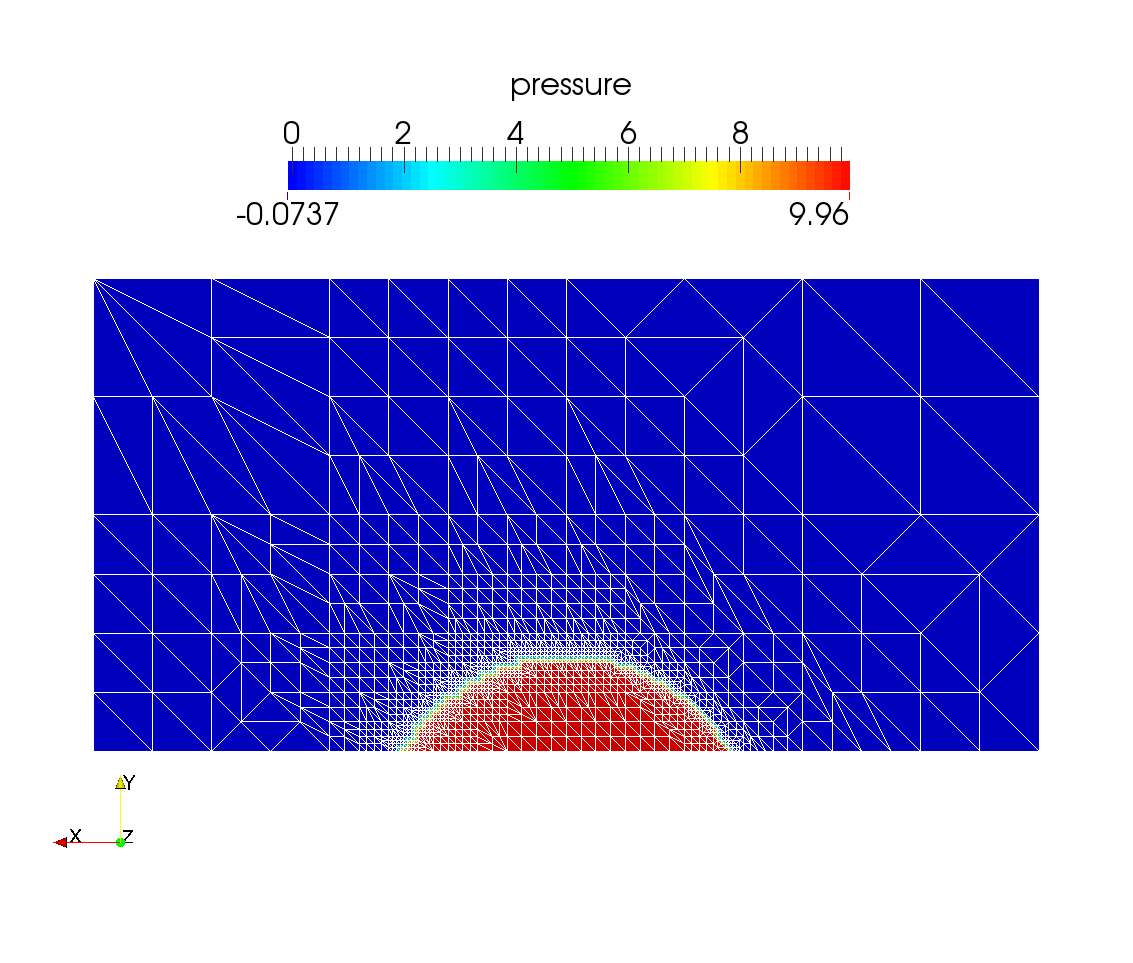}
  \caption{A cross section of the pressure approximation on a locally refined mesh.}
\label{fig:PJumpWithMesh}
\end{center}
\end{figure}

\begin{figure}[ht!]
\begin{center}
  \includegraphics[width=0.85\textwidth,height=0.35\textheight]{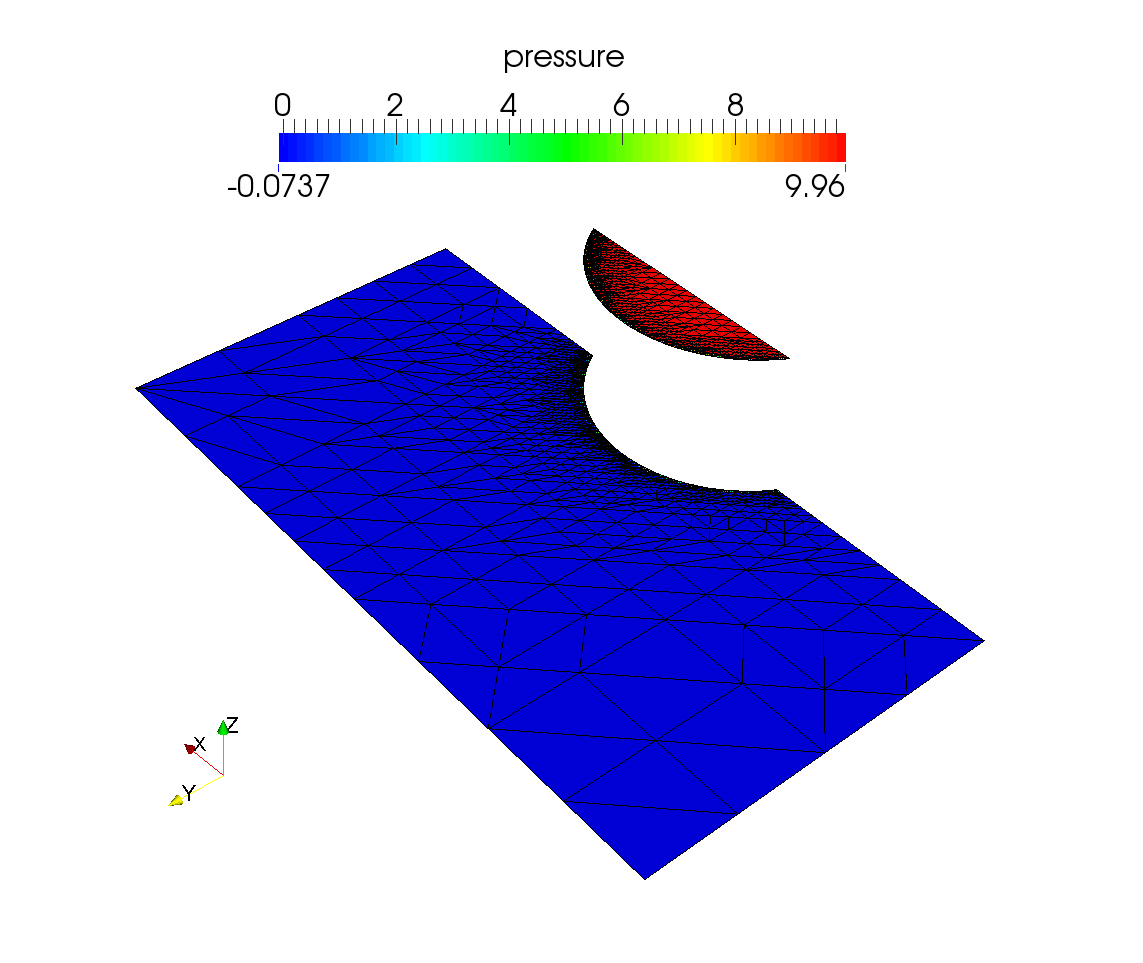}
  \caption{The pressure jump is visualized by scaling in the normal direction of the cross section.}
\label{fig:PJumpVis}
\end{center}
\end{figure}

This problem has the stationary solution
\begin{align*}
 \mbf u^* =0 \text{ in } \Omega_1 \cup \Omega_2, \qquad p^* = \begin{cases} c_0+ k_{\text{const}} &\text{ in } \Omega_1 \\ c_0 &\text{ in } \Omega_2\end{cases}
\end{align*}
where $ k_{\text{const}} = \frac{2\tau}{R} = 10$.  
Due to the discretization of the surface tension forces and the approximation of the pressure solution $p^*$ in the finite element space  spurious velocities will appear. 
The accurate Laplace-Beltrami approach in the discretization \eqref{tt1} and the use of the XFEM method for the pressure approximation
result in  very small spurious velocities. 
In Figure~\ref{fig:PJumpWithMesh} and Figure~\ref{fig:PJumpVis} we illustrate the sharp pressure jump.

We investigate the convergence behavior  of the discrete solution pair $(\bu_h, p_h)$. 
Results are presented in the table~\ref{tab:stabxfem60}. The initial level 0 grid is obtained from a uniform tetrahedral grid that has only 5 vertices (4 intervals) on each edge of the cube $\Omega$, which is then locally refined once close to the interface. Hence, on level 0, the tetrahedra close to the interface have edges of length $0.5*(\frac12)^3= \frac{1}{16}$.   An increase in the level number (first column) in this table corresponds to a local refinement close to the interface of the tetrahedral grid (i.e. one regular local refinement followed by a correction step for elimination of hanging nodes). In Figure~\ref{fig:PJumpWithMesh} a cross section of the level 4 mesh is shown. One important contribution of this paper is the application of the stabilized XFEM method to this problem class. From experiments it follows that the behavior of this method is very satisfactory. In particular, there is no strong sensitivity with respect to the choice of the stabilization parameter $\epsilon_p$ in \eqref{e:decoupled1}. To illustrate this we present results for $\epsilon_p=1$  and for $\epsilon_p = 0.01$. For these two values of $\epsilon_p$ the discrete velocities do not show significant differences and therefore we only show the errors for the case $\epsilon_p=1$. We observe only a mild deterioration of the pressure quality if the stabilization parameter is reduced from  $\epsilon_p=1$ to $\epsilon_p = 0.01$.
% In all further experiments we take $\epsilon_p=1$ {\bf Q: correct? A: It is not the case for next experiment. For next experiment, I have to take very small $epsilon$ to get good result.}.
\begin{table}[ht!]
\begin{center}
  \begin{tabular}{c|c|c|c|c|c|c}
  \hline
  Mesh lvl & $\|e_u\|_{L_2}, \epsilon_p = 1$   & order   & $\|e_p\|_{L_2}, \epsilon_p = 1$  & order   & $\|e_p\|_{L_2}, \epsilon_p = 0.01$ & order   \\
  \hline
  0    & 6.79E-5    & --      & 5.37E-3  & --      &5.31E-3  & --      \\
  \hline
  1    & 1.54E-5    &  2.137  & 1.24E-3    &  2.13  &1.34E-3    & 1.97   \\
  \hline
  2    & 3.18E-6    &   2.281 & 3.38E-4    &   1.87   &3.89E-4   & 1.79  \\
  \hline
  3    & 7.33E-7    &  2.117  & 8.53E-5    &  1.99   & 1.09E-4   &  1.85 \\
  \hline
  4    & 1.69E-7    &  2.120  & 8.53E-5    &  1.99   & 1.09E-4   &  1.85 \\
  \hline
  5    & 4.08E-8    &  2.048   & 5.78E-6  &  1.93   & 1.04E-5   &  1.67  \\
  \hline
  6    & 1.00E-8    &  2.025   & 1.57E-6  &  1.89   & 3.63E-6   &  1.52  \\
  \hline
  \end{tabular}
\end{center}
\caption{Convergence behavior for a static droplet on a plate with $\theta_e=60^\circ$ and different values $\epsilon_p = 1, 10^{-2}$ in the XFEM stabilization.}
 \label{tab:stabxfem60}
\end{table}

An important conclusion is that these results   show a very good agreement with results from a similar experiment \emph{without a contact line}, see subsection 7.10 in \cite{GrossReuskenBook}. Hence, the numerical treatment of the contact line force and of the Navier boundary condition is satisfactory (in this test problem).

The (very small) errors in velocity are caused by spurious velocities close to the interface. The suboptimal order of convergence in velocity is  
caused by the errors in the discretization of the surface tension force  \cite{Grande2015}.
As a further error measure we determine a numerical contact angle $\theta_h$ and compare it to the exact one, $\theta_e$. The numerical contact angle $\theta_h$ is defined as the average of the local (per tetrahedron) approximate angles that are obtained from the  level set function normals \eqref{defntilde}. In a similar way a discrete radius $r_h$, which approximates $r$, is determined. The results are given in 
Table~\ref{tab:ThetaR} and show (approximately) the expected second order convergence.
\begin{table}[ht!]
\begin{center}
  \begin{tabular}{c|c|c|c|c}
  \hline
  Mesh lvl & $|\theta_h - \theta_e|/\theta_e$   & order   & $|r - r_h|/r$ & order   \\
  \hline
  0    & 3.07E-2    & --      &1.44E-2  & --      \\
  \hline
  1    & 1.06E-2    &  1.53  &3.42E-3    & 2.08  \\
  \hline
  2    & 2.46E-3    &   2.11   &8.16E-4   & 2.07  \\
  \hline
  3    & 4.75E-4    &  2.37   & 2.09E-4   &  1.97 \\
  \hline
  4    & 1.31E-4    &  1.85  & 5.19E-5    &  2.01 \\
  \hline
  \end{tabular}
\end{center}
\caption{Errors in approximation of contact angle $\theta$ and radius $r$.}
 \label{tab:ThetaR}
\end{table}
 
\subsubsection*{A  droplet sliding to  equilibrium state}
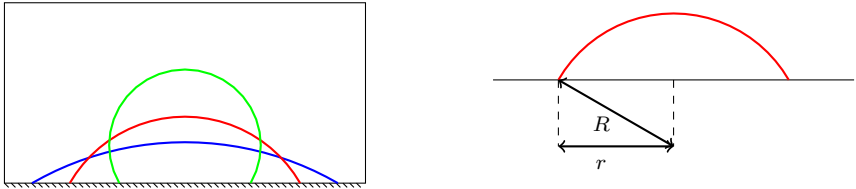
\begin{figure}[ht!]
 \begin{minipage}{0.49\linewidth}
  \centering
  \begin{tikzpicture}[scale =1.2, font=\footnotesize]
    \draw (4,0) -- (0,0) -- (0,2) -- (4,2)--cycle;
    \draw [decoration={border,segment length=0.9mm,amplitude=0.8mm,mirror,angle=45},decorate](0,0) -- (4,0);
    \draw [blue,thick,domain=60:120] plot ({2 + 3.3876*cos(\x)}, {-3.3876 * cos(30) + 3.3876*sin(\x)});
    \draw [red,thick,domain=30:150] plot ({2 + 1.4736*cos(\x)}, {-1.4736 * cos(60)  + 1.4736*sin(\x)});
    \draw [green,thick,domain=-30:210] plot ({2 + 0.83994*cos(\x)}, { -0.83994 *cos(120)+ 0.83994*sin(\x)});
    %\node at (2,0.25) {$\Omega 1$};
    %\node at (2,1.5) {$\Omega 2$};
  \end{tikzpicture}
 \end{minipage}
\begin{minipage}{0.49\linewidth}
  \centering
   \begin{tikzpicture}[scale =1.2, font=\footnotesize]
    \draw (4,0) -- (0,0);
    \draw [red,thick,domain=30:150] plot ({2 + 1.4736*cos(\x)}, {-1.4736 * cos(60)  + 1.4736*sin(\x)});
    \draw [<->, thick] (2, {-1.4736 * cos(60)}) -- ({2 + 1.4736*cos(150)}, {-1.4736 * cos(60)  + 1.4736*sin(150)});
    \draw [<->, thick] (2, {-1.4736 * cos(60)}) -- ({2 + 1.4736*cos(150)}, {-1.4736 * cos(60)} );
    \draw [dashed] (2, 0) -- (2, {-1.4736 * cos(60)});
    \draw [dashed] ({2 + 1.4736*cos(150)}, 0) -- ({2 + 1.4736*cos(150)}, {-1.4736 * cos(60)});
    \node at (1.2, {-1.4736 * cos(60) + 0.25}){$R$};
    \node at (1.2, {-1.4736 * cos(60) - 0.20}){$r$};
  \end{tikzpicture}
 \end{minipage}
 \caption{2D sketch of the static droplet  shape and of the radii $R$ and $r$.}
 \label{fig:Rr}
\end{figure}

We extend the previous test case to a non-stationary problem.
We take the same setup as described above except that  we take the initial contact angle $\theta_0 = 90^\circ$ and take different equilibrium contact angles $\theta_e \neq \theta_0$.
We consider the three cases $\theta_e = 30^\circ, \theta_e = 60^\circ \text{ and } \theta_e = 120^\circ$, cf.~Figure~\ref{fig:Rr}.
When we release the droplet with  initial contact angle $\theta_0$ it will  spread or contract to a spherical cap with  contact angle $\theta_e$.

The time dependent solution of this problem converges to  a static state, which is known (essentially the same as in the static droplet case). 
Note that in this test case we have an evolving interface (and contact line) and thus the level set method and its coupling with the Navier-Stokes equations play a key role. 

 By using conservation of  volume of the droplet, we can calculate the radius of the droplet $R$ and the radius $r$ of the circle inside the contact line at the static state.
For the 3D case we thus obtain, with $R_0$ the droplet radius at $t=0$:
\begin{equation}
 R = R_0 \sqrt[3]{\frac{2}{ 2- 2\cos \theta_e -\sin^2 \theta_e  \cos \theta_e} }, \quad r = R\sin \theta_e.
\label{eq:Rr}
\end{equation}
(In \cite{dupont2010} the 2D analogon of this formula is given). 

%\begin{figure}[ht]
%\centering
% \includegraphics[trim=2.5cm 6cm 2.5cm 6cm, clip=true, totalheight=0.5\textheight]{radiusheight.pdf}
%\caption{2D abstract plot for the static state shape and the notations for $R$ and $r$.}
%\end{figure}
We perform a simulation until the equilibrium is reached (approximately).  For the numerical equilibrium solution we determine a numerical contact angle $\theta_h$ and discrete radius in the same way as in the static droplet case above. The corresponding (relative) errors on mesh level 1, which has the same resolution as the level 1 mesh in the previous experiment, are shown in table~\ref{tab:releasingdroplet}. We note that in this experiment, due to the dynamics of the interface, the locally refined grid is changing in time during the simulation   (local grid refinement and coarsening).
\begin{table}[ht!]
\begin{center}
  \begin{tabular}{c|c|c|c|c|c}
  \hline
  $\theta_e$        & $\theta_{h}$   &  $|\theta_h - \theta_e|/\theta_e$& $r$        &  $r_h$ &  $|r_h - r|/r$\\
  \hline
  30$^\circ$        & 29.77$^\circ$   &     7.7E-3      &  0.169384    &0.169425  & 2.42E-4 \\
  \hline
  60$^\circ$        & 59.74$^\circ$   &     4.3E-3     &  0.127619    &0.127408  &  1.65E-3\\
  \hline
  120$^\circ$       & 119.6$^\circ$   &     3.3E-3     &  0.0727416  &0.0733349  &  8.15E-3\\
  \hline
  \end{tabular}
\end{center}
\caption{Errors in contact angle $\theta_e$ and radius $r$ for three cases on the level 1 mesh.}
 \label{tab:releasingdroplet}
\end{table}

Comparing the results in Table~\ref{tab:releasingdroplet} with the level 1 results in  Table~\ref{tab:ThetaR}, we  see that these have the same order of magnitude.  The errors in velocity and pressure  for the numerical equilibrium solution on mesh level 1 are $\|e_u\|_{L^2}=$2.03E-6 and $\|e_p\|_{L^2}=$1.13E-3. These are  of the same order of magnitude (or even smaller) as the level 1 errors in Table~\ref{tab:stabxfem60}.  From this we conclude that these results for the time dependent simulation are very satisfactory. 
Under mesh refinement we observe convergence which, however, does not show a regular second order behavior. This is caused by an undesirable behavior of the reparametrization and volume correction method, which is not addressed in this paper.  
%\begin{figure}[ht]
%\centering
% \includegraphics[scale=0.5]{Spread60Err.png}
%\caption{The error at quasistatic state for a spreading droplet}
%\label{fig:spread60}
%\end{figure}
%
%We also measured the $L_2$ norm error of velocity and pressure at the static state with respect to mesh size for the case $\theta_e =60^\circ$. 
%In figure~\ref{fig:spread60}, we show the results in 3 meshes where the first mesh index is the initial number of intervals and the second mesh index is the number of adaptive %refinement near the interface. 
%At t =1.6, the spreading droplet almost reaches its static state and the error measured for both velocity and pressure stays the same. 
%At t=2.5, the volume correction is turned off. The error of velocity and pressure in finest mesh decreased drastically. 
%Our current simple volume correction strategy has big (negative) impact to the solutions of fine meshes. 
%It is also related to that our analytical solution is also strongly dependent on interface position.
%When we keep using our volume correction, we have convergence problem caused by volume correction.
%When we turn off the volume correction, there are two error sources. One is the error from our general numerical methods, the other is volume change.
%Our results still converge to the analytical solution without volume change dominates the error.
%When we continue refine meshes, the volume correction error will dominate the whole error, we will lose our (optimal) convergence order as the stationary test case.

\subsection{Couette flow} 
In this section we do a further validation of the solver by considering a  two-phase  Couette flow in a three-dimensional channel $[0,L_x]\times [0,L_y] \times [0,L_z]$
with moving contact lines. In the first experiment, cf.~Figure~\ref{fig:CouetteFlow}, the two fluids are separated by two different interfaces, whereas  in the second experiment, cf.~Figure~\ref{fig:couetteflow2}, there is only one interface between the fluids. 
The upper and lower boundaries are walls that move  with velocity $U_{wall}$ and $-U_{wall}$ in $x$-direction, respectively.

We show results of two types of numerical experiments. First we present results obtained with our solver and study grid convergence by using a reference solution that is obtained by applying the  solver on a very fine grid. In the second experiment we compare the simulation results of our solver with molecular dynamics simulation results from the literature.   
 In both cases, we set $\theta_0 = \theta_e = 90^\circ$.

\subsubsection*{Convergence study with a fine mesh reference solution} 
\begin{figure}[ht!]
\begin{center}
  \begin{tikzpicture}[scale=1.9, font=\footnotesize]
    %\hspace*{0.8cm}
    \draw [thick,->](-0.5,-0.5) -- (-0.5,-0.1);
    \draw [thick,->](-0.5,-0.5) -- (-0.1,-0.5);
    \node at (-0.5,0) {$z$};
    \node at (0,-0.5) {$x$};
    \draw [thick,->] (0,1.5) -- (5,1.5);
    \draw [thick,->] (5,0) -- (0,0);
    \draw (1.75,0) -- (1.75,1.5);
    \draw (3.25,0) -- (3.25,1.5);
    \draw (0,0) -- (0,1.5);
    \draw (5,1.5) -- (5,0);
    \draw (3.25,0) -- (3.25,1.5);
    \node at (-0.1,-0.1){0};
    \node at (5,-0.1) {$L_x$};
    \node at (-0.1,1.5) {$L_z$};
    \node at (1.0,0.75) {Fluid 1};
    \node at (2.5,0.75) {Fluid 2};
    \node at (4.0,0.75) {Fluid 1};
    \node at (2.5,-0.1) {$-U_{wall}$};
    \node at (2.5, 1.6) {$U_{wall}$};
  \end{tikzpicture}
  \caption{Sketch of a cross section of the Couette two-phase channel flow.}
\label{fig:CouetteFlow}
\end{center}
\end{figure}
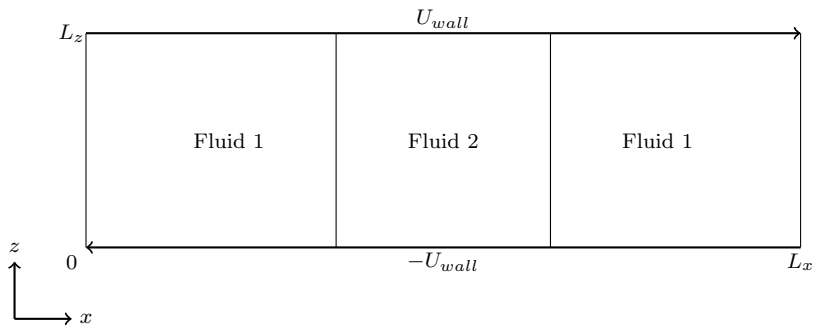

The setup is as described above, with a Navier boundary condition on the upper and lower (moving) boundaries, a periodic boundary condition on left and right boundaries and a symmetry boundary condition in $y$ direction. We take length $L_x=1$, width $L_y=0.2$ and hight $L_z=0.4$. The initial planar interfaces  are as sketched in Figure~\ref{fig:CouetteFlow}.

 The initial velocity is  $\mbf u=0$ . 
Densities and viscosities are taken as $\rho_1=\mu_1=3$,  $\rho_2=\mu_2=1$.  The slip coefficient in the effective wall force $\mbf f_S$ is set to $\beta_S=1.2$   
 and the surface tension coefficient is $\tau =5.5$. These two parameter values are the same as in the MD simulations from \cite{Qian03}, which are considered in the next experiment.
In this test, we take an initial uniform tetrahedral triangulation with 5 vertices (4 intervals) on the edges of the channel in $x$-direction, 3 vertices (2 intervals) on the edges in $z$-direction an 2 vertices (1 interval) on the edges in $y$-direction. We use uniform refinement. The (technical) reason for this is that with uniform refinement the interpolation of coarse grid functions to finer grids is obvious and there are no interpolation errors. For $t=0.032 $ we compute a reference solution $ \bu^{Ref}$ on the tetrahedral triangulation obtained after 5 refinements of the initial one. We use such a short time interval $[0, 0.032]$  due to computational limitations for computing the reference solution on the five times refined triangulation. For $t=0.032$ the velocity solution as a size $ \|\bu^{Ref}\|_{L^2}= $8.76E-2. The velocity errors on coarser grids (for $t=0.032$)  are shown in Table~\ref{tab:couette}.
% \begin{table}[ht!]
% \begin{center}
%   \begin{tabular}{c|c|c}
%   \hline
%   $Ref.$ & $\|\bu^{Ref}-\bu_h\|_{L_2}$   &   order     \\
%   \hline
%   0        & 9.1e-3             &  -      \\
%   \hline
%   1        & 2.8e-3             &  1.7    \\
%   \hline
%   2        & 1.4e-3             &  1.0    \\
%   \hline
%   3        & 4.8e-4            &  1.5     \\
%   \hline
%   \end{tabular}
% \end{center}
% \caption{The convergence behavior of velocity.}
%  \label{tab:couette}
% \end{table}
\begin{table}[ht!]
\begin{center}
  \begin{tabular}{c|c|c}
  \hline
  $Ref.$ & $\|\bu^{Ref}-\bu_h\|_{L_2}$   &   order     \\
  \hline
  0        & 8.6e-3             &  -     \\
  \hline
  1        & 2.5e-3             &  1.8    \\
  \hline
  2        & 1.3e-3             &  0.94   \\
  \hline
  3        & 4.4e-4             &   1.6   \\
  \hline
  \end{tabular}
\end{center}
\caption{The convergence behavior of velocity.}
 \label{tab:couette}
\end{table}

\begin{figure}[htb!]
\centering 
\includegraphics[scale=0.6]{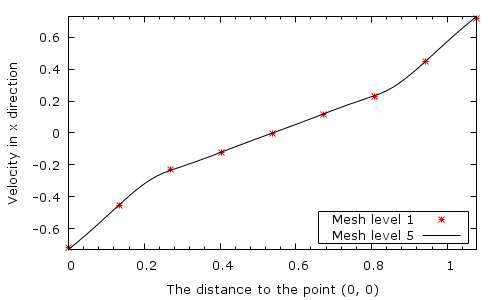}
\caption{The velocity in $x$ direction from the point $(0, 0)$ to the point $(1, 0.4)$ in $x$-$z$ plane at t=0.032.}
\label{fig:LinePlotCouette}
\end{figure}

\begin{figure}[htb!]
\centering
\includegraphics[trim=0cm 6.5cm 1.5cm 1.5cm, scale=0.26]{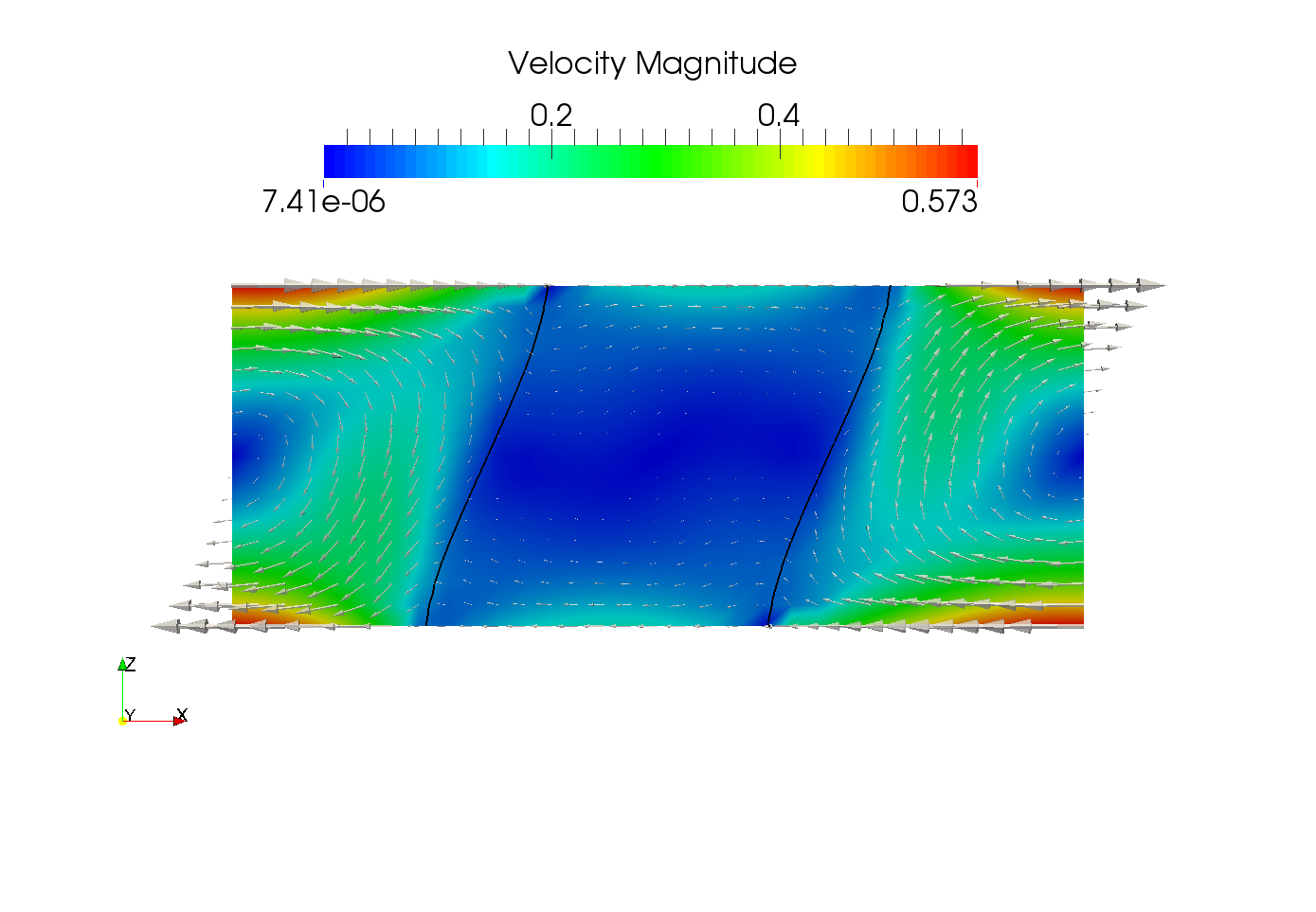} 
\caption{The couette flow at t=0.32, 2 times refined triangulation.}
\label{fig:couetteflow1}
\end{figure}

We observe that on  average we get a close to 1.5 order of  convergence for the velocity. Furthermore,  note that after one refinement we have a relatively coarse mesh but already a rather accurate numerical solution. To illustrate this, in Figure~\ref{fig:LinePlotCouette}, we show a line plot of the $x$-velocity in $x$-$z$ plane at $y=0.1$ from the point $(0, 0)$ to the point $(1, 0.4)$ for the reference mesh and the level 1 mesh. 
% In Figure~\ref{fig:LinePlotCouetteCorner}, a zoomed in line plot at the lower left corner of the domain is showed. These two pictures demonstrate the solutions of coarse meshes converge to the solution in the reference mesh. 
In Figure~\ref{fig:couetteflow1} and \ref{fig:couetteflow2}, we show a cross section at $y=0.1$ of the computed solution on a 2 times refined triangulation at $t=0.032$ and $t=0.32$, respectively.

%\begin{figure}[htb!]
%\centering 
%\includegraphics[scale=0.5]{LinePlotCouette2.png}
%\caption{The velocity in x direction from the point $(0, 0)$ to the point $(0.75, 0.3)$ in x-z plane at t=0.032 }
%\label{fig:LinePlotCouetteCorner}
%\end{figure}
%In Figure~\ref{fig:couetteflow1} we show a cross section at $y=0.1$ of the computed solution at $t=0.32$ of the 2 times refined triangulation.\\[2ex]
\begin{figure}[htb!]
\centering 
\includegraphics[trim=0cm 6.5cm 1.5cm 1.5cm, scale=0.26]{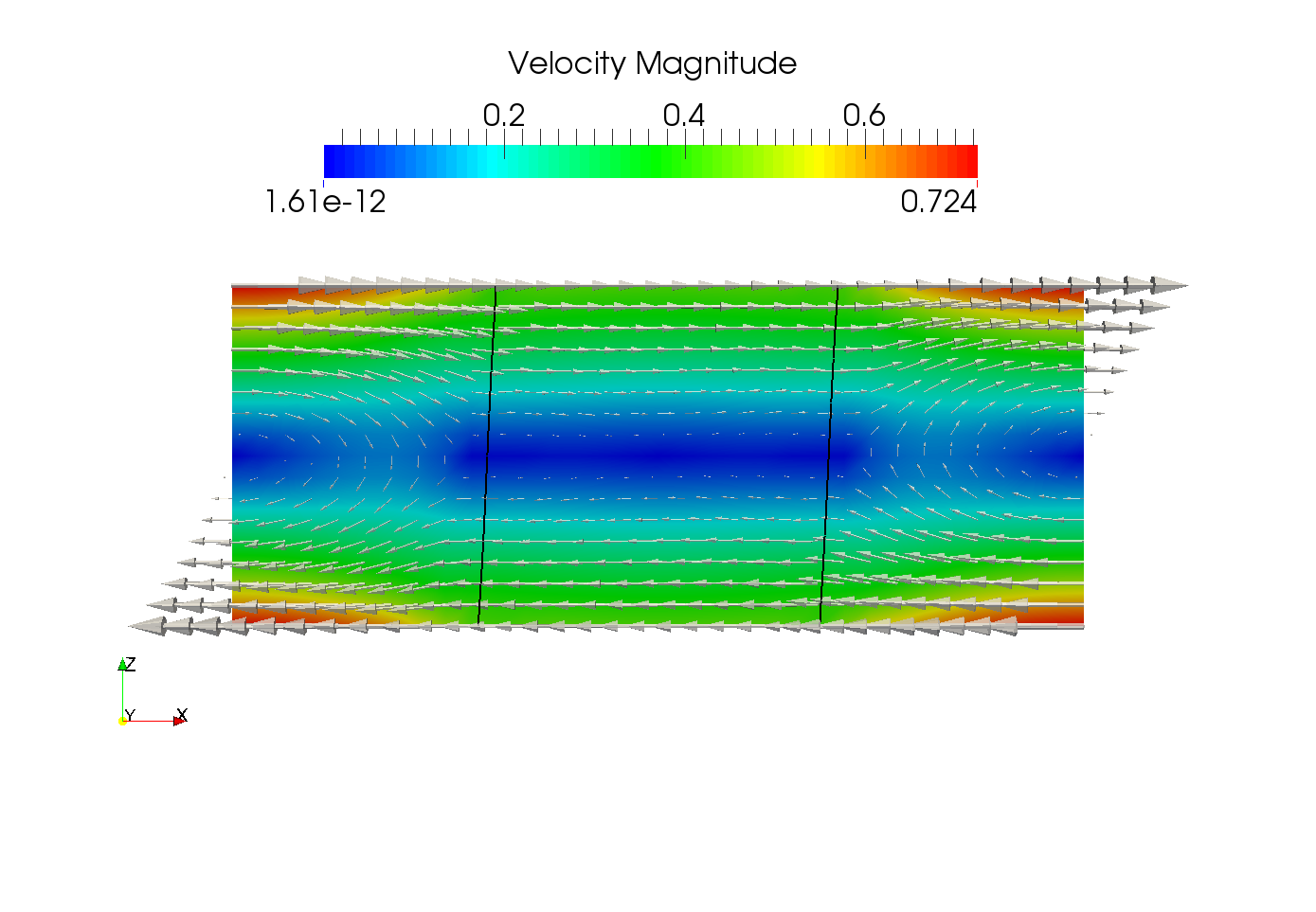}
\caption{The couette flow at t=0.032, 2 times refined triangulation.}
\label{fig:couetteflow0}
\end{figure}

\subsubsection*{Comparison with molecular dynamics simulation} 
We performed a second numerical simulation experiment for the Couette flow in which  
results are compared with MD simulations in \cite{Qian03}.  

\begin{figure}[ht!]
\centering 
\includegraphics[width=0.8\textwidth]{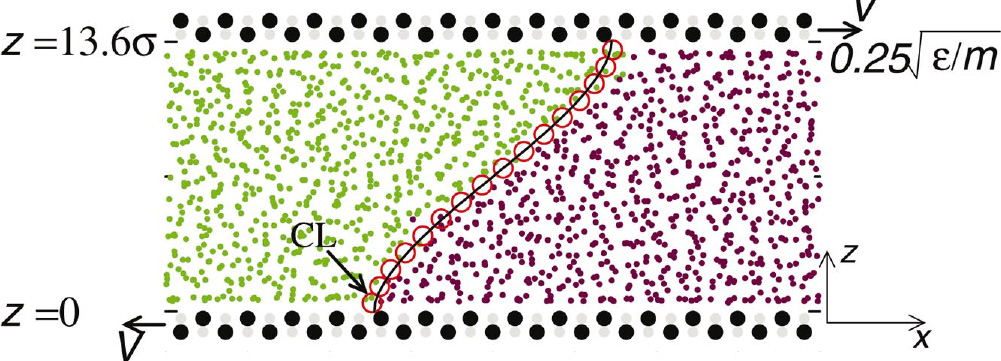}
\caption{The MD Couette flow setup in  \cite{Qian03}.}
\label{fig:couetteflow2}
\end{figure}

The MD setup from \cite{Qian03} is shown  in 
Figure~\ref{fig:couetteflow2}. It is a 3D channel 
with length $L_x=163.5\sigma$, width $L_y=6.8\sigma$ and hight $L_z=13.6\sigma$,
where $\sigma$ is the relevant unit (a microscopic length scale in the MD simulation). 
Note that opposite to the previous experiment, there is \emph{one} interface between two immiscible fluids  in the middle of the channel. . 
The density of the two fluids are $\rho_1=\rho_2=0.81m/\sigma^3$ and their viscosities are 
$\mu_1=\mu_2=1.95\sqrt{m\epsilon}/\sigma^2$, where $m$ and $\epsilon$ are suitable mass 
and energy scales in the MD simulation. From now on we delete the units.  The slip coefficient in the effective wall force is set to 
$\beta_S=1.2$. The surface tension is $\tau = 5.5$.  The wall velocity, cf. Figure~\ref{fig:CouetteFlow}, is $U_{wall}=0.25$.
 In this system a  steady state is reached as illustrated  in Fig.~\ref{fig:couetteflow2}. 

The MD simulations provide  velocity  profiles, which will be compared with the numerical simulations by our finite element method.
In our simulations we take the channel given by $(-L_x/2,L_x/2)\times(0,L_y)\times(0,L_z)$, and we use a periodic boundary condition in the $y$-direction.
On the left and right boundary, we take Dirichlet boundary condition given by  $\mbf u(y,z)=\frac{2U_{wall}(z-L_z/2)}{L_z+2*l_{s}}\mathbf{e}_1$, for $(y,z)  \in (0,L_y) \times (0,L_z)$, where
$l_{s}=\mu/\beta_S$ is the slip length. 
 In this test, we take an initial tetrahedral triangulation with 11 vertices (10 intervals) on the edges of the channel in $x$-direction, 5 vertices (4 intervals) on the edges in $z$-direction an 3 vertices (2 interval) on the edges in $y$-direction, and we apply 3 adaptive refinements near the interface.
%All the parameters are determined by the MD simulations.

\begin{figure}[ht!]
\centering 
{\includegraphics[width=120mm]{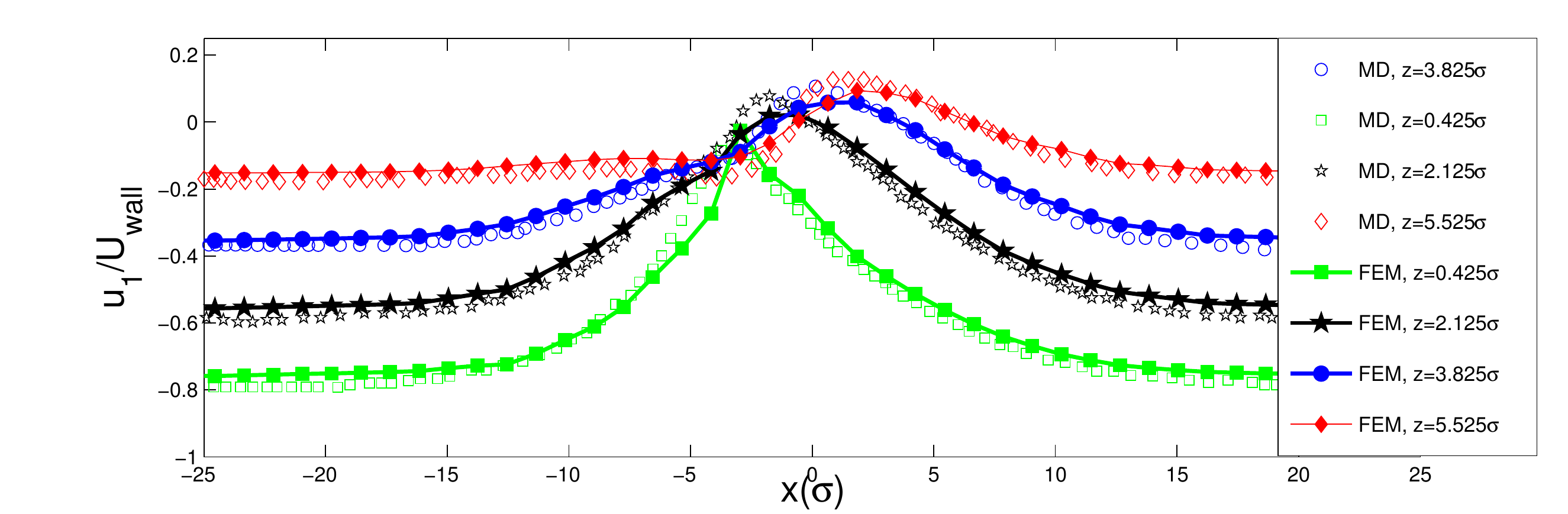} }
\caption{Comparisons of MD simulations and the sharp-interface model simulations.}
\label{fig:CompareMD}
\end{figure}

Numerical results for the computed equilibrium solution are shown in Fig.~\ref{fig:CompareMD}. We show the profile of the relative velocity in $x$-direction
 $u_1/U_{wall}$, where $u_1$ denotes the $x$-component of velocity. The profiles are shown for $u_1(x,\frac12 L_y, z_j)$, with $x \in [-25,25]$, $z_1=0.425$ (green/bottom), $z_2=2.125$ (black), $z_3=3.825$ (blue) and $z_4=5.525$ (red/top). 
The empty symbols show the results from MD simulations and the filled symbols with solid lines show the results by our 
FEM method. We observe a good agreement between the two sets of simulation results. Note that differences may be due not only to numerics but also to modeling discrepancies. Near the wall boundary, at $z_1=0.425$ (bottom curve),
the $x$-velocity $u_1(x,\frac12L_y, z_1)/U_{wall}$ is close zero for $x$ close to the contact line, which implies that the fluids slip there (note that the wall moves).

\subsection{Wetting on a curved boundary} The following example demonstrates how our method performs for a contact line problem with a curved contact boundary. 
%It is well-known that using standard polygonal approximation for a curved domain with strong imposition of zero normal velocity at the boundary will leads to zero velocity at the corners \cite{verfurth1986,urquiza2014weak}. %See more discussions in Remark~\ref{rem:babuskaParadox}.
We solve a sliding droplet problem in a hemispherical domain. 
A droplet is set at the bottom of the hemisphere with a initial contact angle $\theta_0 = 104.5^\circ$.  
The rest of the domain is filled with another fluid.
The density of the droplet is set to $\rho_1 = 3$ and the viscosity is $\mu_1 = 3$.  The fluid outside has $\rho_2=1$ and $\mu_2 =1$.  We ignore the effect of the gravity.
The initial velocity is zero. We set the equilibrium contact angle $\theta_e=60^\circ$. 
Due to the difference between $\theta_0 $ and $\theta_e $, the droplet will spread on the bottom curved boundary and it is expected to reach a static state with a contact angle $\theta_d = \theta_e$. 
The spreading process and the static state are similar to the sliding droplet experiment in subsection 9.1. The only difference is that the contact boundary is curved.
 
\begin{figure}[htb!]
\centering 
\includegraphics[scale=0.27]{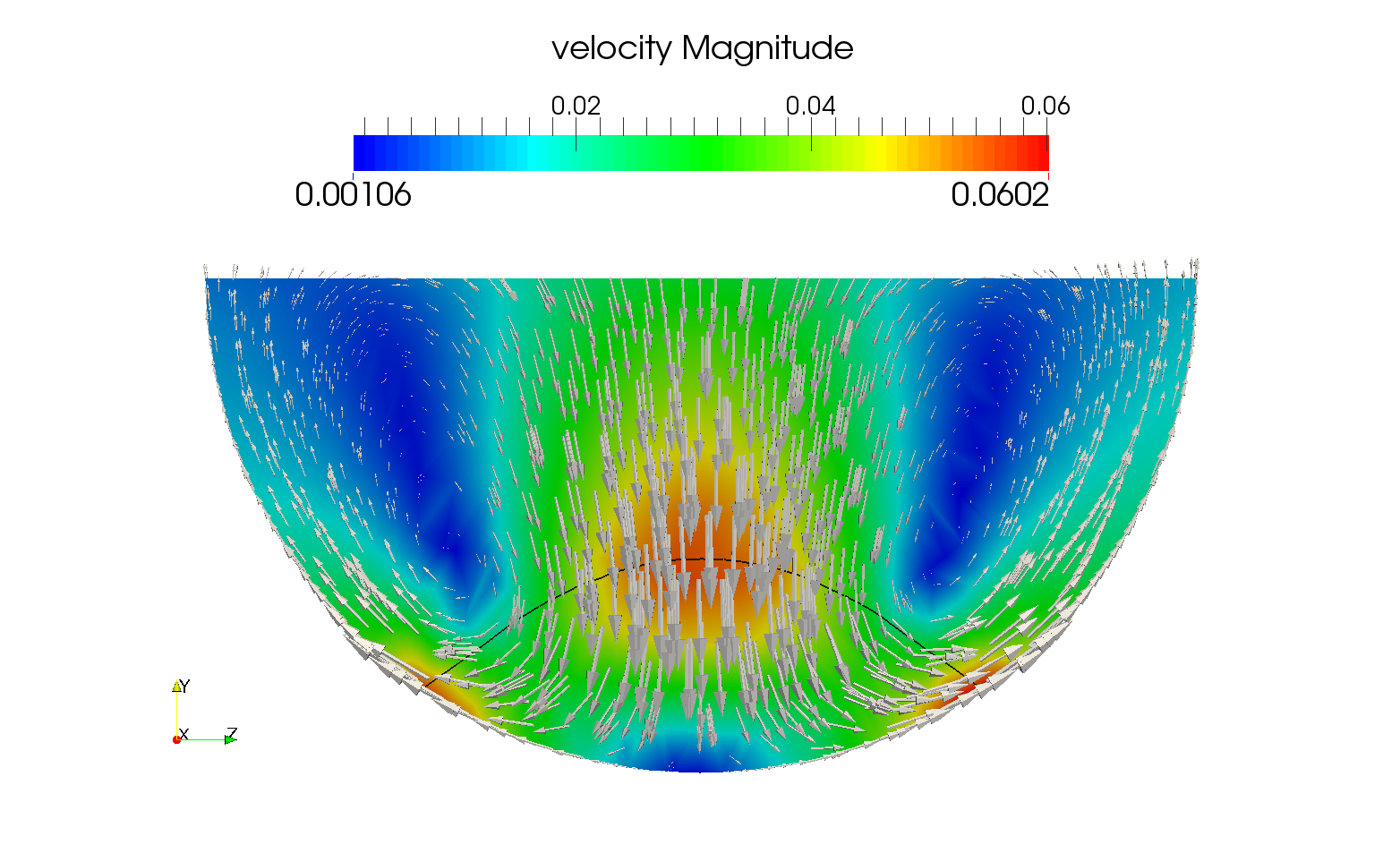}
\caption{Wetting on a curved boundary.}
\label{fig:curvebnd}
\end{figure}

In Figure~\ref{fig:curvebnd} we show a result of the dynamics of the droplet for $t=0.2$.  In this simulation we use uniform grids with 16 intervals in each direction. The contour color denotes the velocity magnitude, and the white arrows represent the flow field.  
At the beginning stage $t=0.2$,  a significant velocity, tangential to the curved boundary, appears in the vicinity of the contact lines. 
From the cross section result, we also observe that  two vortices are formed.  
%To keep volume conservation, the velocity is also large on the top of the droplet. 
For larger $t$  ($t\approx 5$) the droplet becomes stationary and the contact angle is almost $60^\circ$. 
Table~\ref{tab:curved} shows  results for the numerical contact angle and contact area at the static state.
The experiment shows that Nitsche's method can handle the Navier boundary condition on the curved boundary and results in an accurate approximation of the static state of a sliding droplet. In Nitsche's method, for $\bn_S$ we used the normal to the piecewise planar approximation (faces of the tetrahedra) of the curved domain. We are aware of the fact that this choice for the normal may lead to a suboptimal discretization scheme, cf.~\cite{urquiza2014weak,Dione2015}. This is related to Babuska's paradox \cite{verfurth1986}: the solution of stationary Stokes equations with slip boundary conditions on polygonal domains approximating the smooth domain do not converge to the solution of the smooth limit domain. A systematic investigation of this topic is left for future research.  
%This example shows that our method avoids the Babuska's paradox\cite{verfurth1986,urquiza2014weak}, which implies the zero velocity on the curved boundary.
\begin{table}[ht!]
\begin{center}
  \begin{tabular}{c|c|c|c}
  \hline
  $\theta_e$ & $\theta_{e,h}$   &   $A$          &  $A_h$ \\
  \hline
  60$^\circ$        & 60.35$^\circ$  & 0.055     &  0.054 \\
  \hline
  \end{tabular}
\end{center}
\caption{Comparison of analytical and discrete contact angle $\theta_e$ and contact area $A$.}
 \label{tab:curved}
\end{table}

%\subsection{falling droplets on inclined plane}
%  \begin{figure}[ht!]
%\vspace*{0mm}
%    \centering
%  \resizebox{!}{7cm}%{height=6.5cm,width=10cm}
%    {\includegraphics[ height=60mm]{wettingCurveBnd0001.jpg}}
%    %\vspace*{-10mm}
%    \caption{A liquid droplet on a curved domain.}
%   \lbl{fig:curveBnd}
% \end{figure}
%   \begin{figure}[ht!]
%\vspace*{0mm}
%    \centering
%  \resizebox{!}{7cm}%{height=6.5cm,width=10cm}
%    {\includegraphics[ height=60mm]{wettingCurveBnd0060.jpg}}
%    %\vspace*{-10mm}
%    \caption{A liquid droplet on a curved domain.}
%   \lbl{fig:curveBnd1}
% \end{figure}
%\begin{remark}\label{rem:babuskaParadox}
%We are aware that using a polygonal domain to approximate the curved domain one can run into Babuska's like paradox. 
%Babuska's paradox states that: the solution of certain equations (Kirchhoff plate equations) with simple support boundary in a disk is not the limit of the solutions to the same equation posed on polygonal domains approaching the disk.
%We refer to the literature \cite{urquiza2014weak} for methods to remedy Babuska`s paradox in a Stokes's problem.
%The accuracy of our method for the non-static state is still ongoing research for two-phase problems with contact lines.
%\end{remark}

\subsection{Wetting on chemically boundary}
In the last example, we present  numerical results for a liquid droplet sliding on a chemically patterned surface. 
In this problem, due to the inhomogeneity of the solid surface, we have a more challenging contact line dynamics.  
A real three dimensional simulation is required to capture all the characteristics of the hydrodynamics.

In our example,  the bottom  $x-z$ surface is not homogeneous with a spatially varying static contact angle $\theta_e(x, z)$. In  polar coordinates $(x,z)=(r\cos\phi, r\sin\phi)$ the prescribed static contact angles is 
\begin{equation}
\theta_e(r,\phi)=
\left\{\begin{array}{ll}
\frac{\pi}{3}  &  \hbox{if } \phi\in(0,\frac{\pi}{3})\cup(\frac{2\pi}{3},\pi)\cup(\frac{4\pi}{3},\frac{5\pi}{3}),\\
\frac{2\pi}{3} & \hbox{otherwise.}
\end{array}
\right.
\end{equation}
In this setting, the surface contains both hydrophilic ($\theta_e < \frac{\pi}{2}$) and hydrophobic ($\theta_e > \frac{\pi}{2}$) parts.

\begin{figure}[htb!]
\centering 
\subfigure[$t=0$]{\includegraphics[width=61mm]{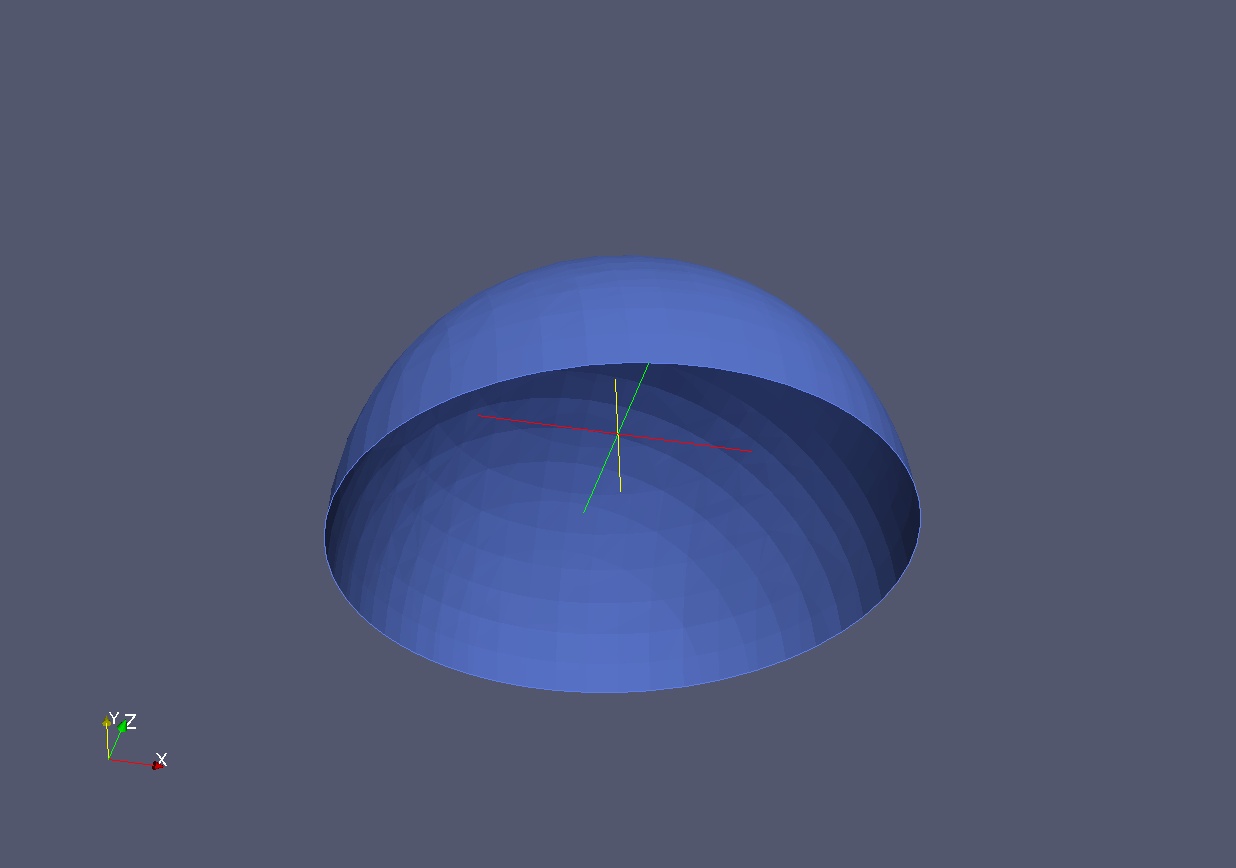}}
\subfigure[$t=0.1$]{\includegraphics[width=61mm]{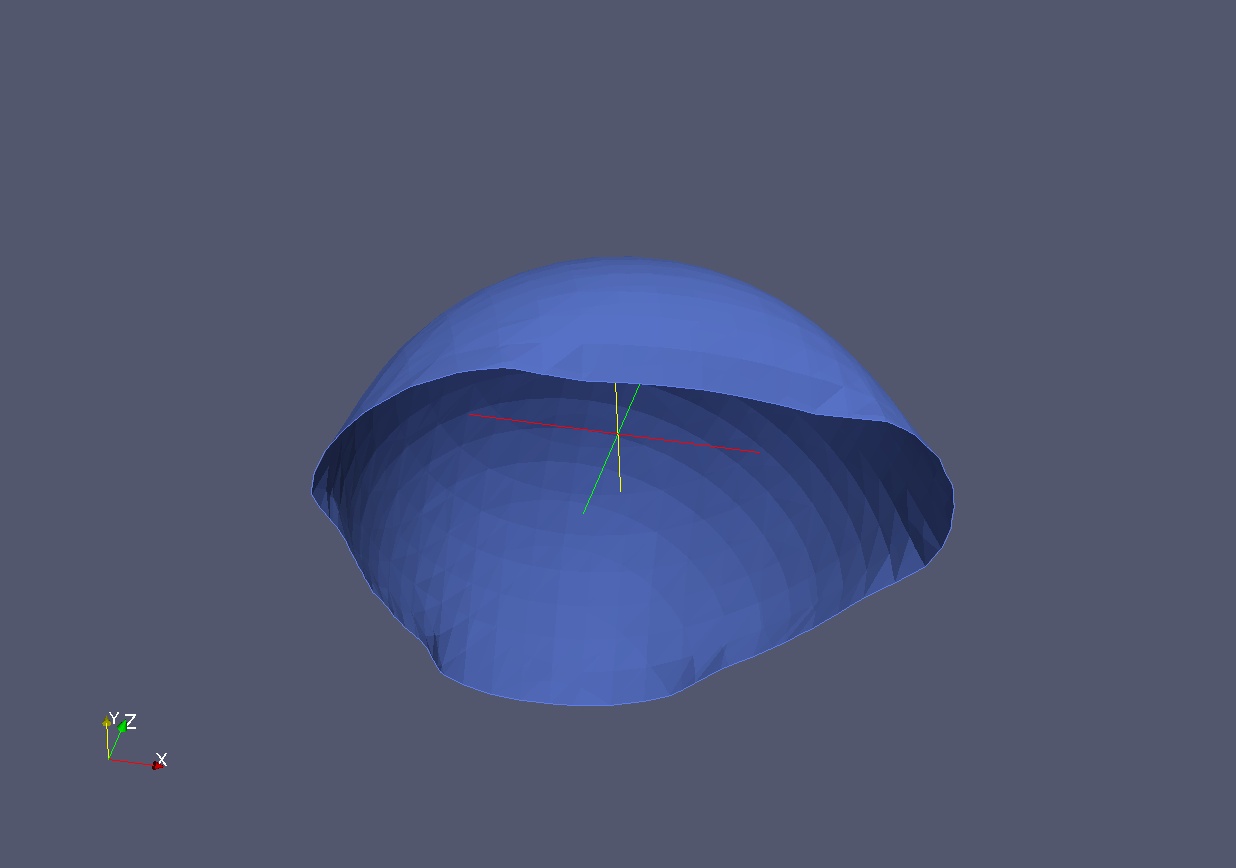}} 
\subfigure[$t=0.2$]{\includegraphics[width=61mm]{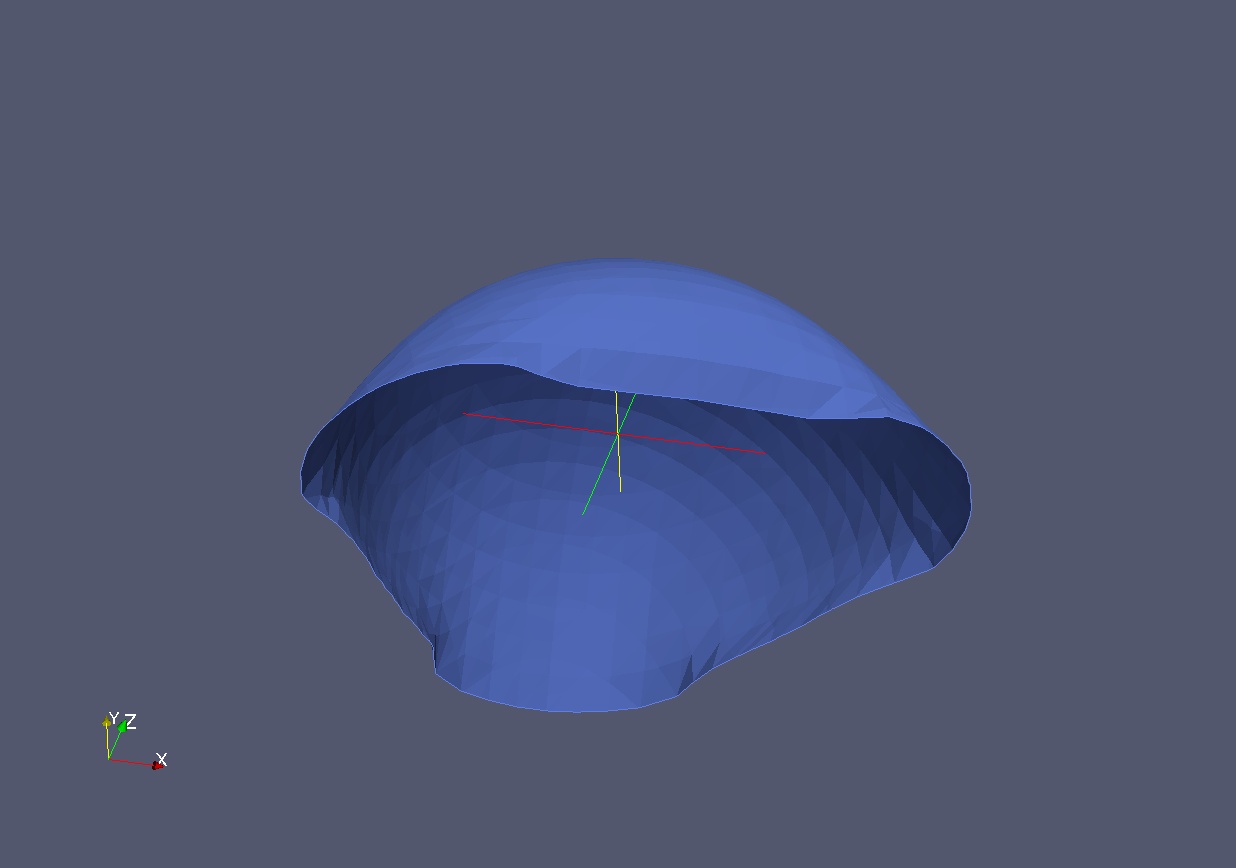}}
\subfigure[$t=0.3$]{\includegraphics[width=61mm]{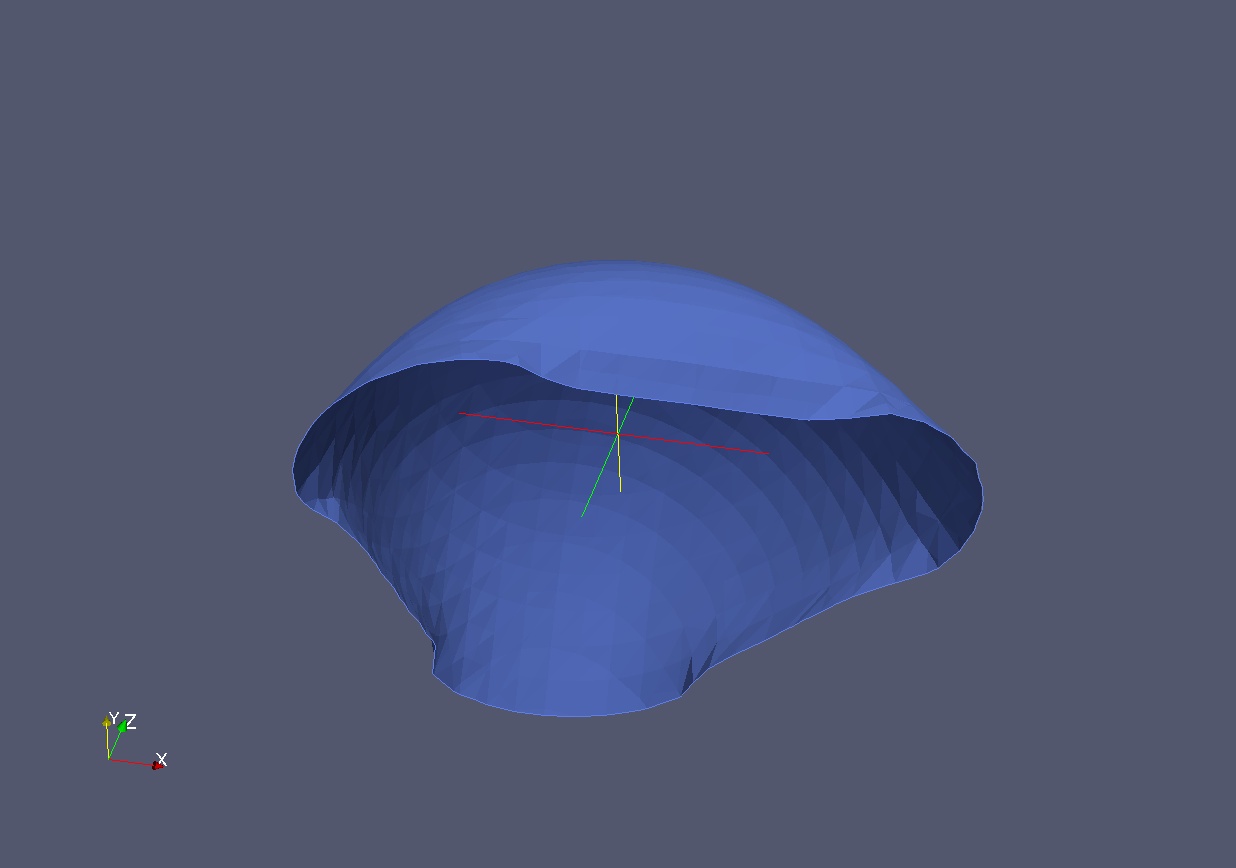}}
\subfigure[$t=0.6$]{\includegraphics[width=61mm]{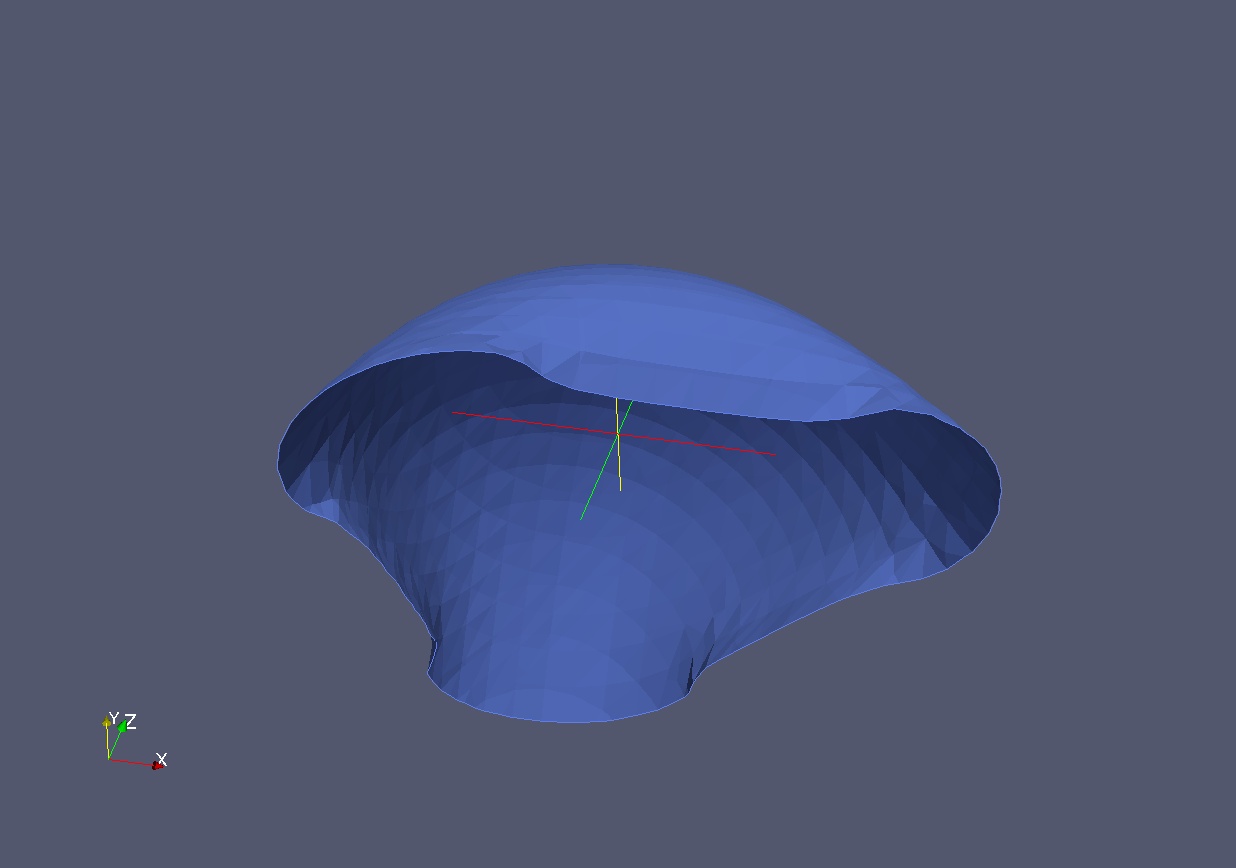}}
\subfigure[$t=1.2$]{\includegraphics[width=61mm]{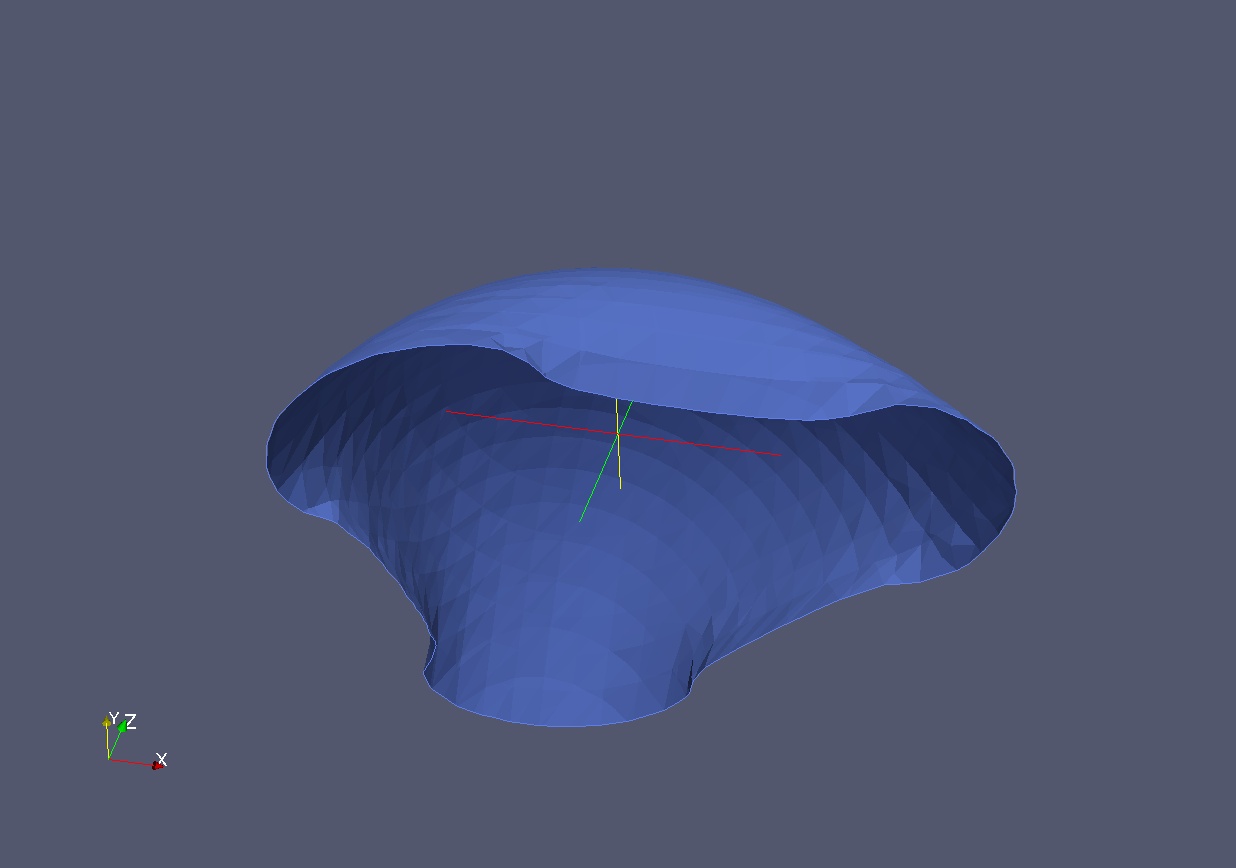}} 
\caption{Wetting on a chemically patterned surface.}
\label{fig:patterned}
\end{figure}

A spherical droplet is placed at the center of the bottom plane.
All other conditions are the same as in  subsection 9.1.
Figure~\ref{fig:patterned} shows some typical shapes of the droplet in the process of wetting.
Starting with a uniform contact angle $\theta=\frac{\pi}{2}$ 
 the droplet moves on the surface and changes its shape. 
It spreads on the hydrophilic part of the bottom and shrinks on the hydrophobic part. 
At $t=1.2$ the droplet is almost stationary. 
The contact angle then is approximately $60^o$ on the hydrophilic part of the surface and approximately $120^o$ on the hydrophobic part, away from  boundaries between the two parts.
The discrete contact angles, which are average values on a segment of the contact line away from the boundaries between the hydrophilic and  hydrophobic parts, are shown in Table~\ref{tab:patterned} with the same resolution as the level 1 mesh in  subsection 9.1. The results in this experiment show that our finite element approach is robust in the sense that in can also deal with this more complicated moving contact line flow problem.
\\
 \begin{table}[ht!]
\begin{center}
  \begin{tabular}{c|c|c|c}
  \hline
  $\theta_A$ & $\theta_{A,h}$   &    $\theta_B$           &  $\theta_{B,h}$ \\
  \hline
  60$^\circ$        & 60.36$^\circ$  & 120$^\circ$     & 119.32$^\circ$ \\
  \hline
  \end{tabular}
\end{center}
\caption{Comparison of discrete contact angles and the prescribed Young's angles in patterned surface. $\theta_A$ denotes the hydrophilic angle and $\theta_B$ denotes the hydrophobic angle.}
 \label{tab:patterned}
\end{table}

\section{Conclusion and outlook} \label{sectoutlook}
We presented finite element techniques for the discretization of two-phase flow problems with moving contact lines. For describing these flow problems a class of models is given and a corresponding variational formulation, which forms the basis for the finite element method, is derived.  The stabilized XFEM method, recently introduced in \cite{Hansbo14} for a stationary two-phase Stokes problem, is also very suitable for two-phase flow problems with moving contact lines. The Nitsche method is very suitable for handling the Navier boundary condition in a convenient and accurate way. We presented a unified approach for discretization of different interface stress tensors $\bsigma_\Gamma$ and contact line forces $\fL$. We consider these techniques to be promising for this problem class. A  further study of these methods by means of theoretical and numerical analyses is planned. In particular the application to specific MCL models (e.g. GNBC) and a systematic comparison with other finite element techniques is a topic of further 
research. 
%\noindent {\bf Acknowledgments.}
%We thank DROPS.\dfrac{\cite}{}{DROPS}
\bibliographystyle{siam}
\bibliography{literatur}
\end{document}